\documentstyle[12pt,fleqn,rotate,epsfig]{article}
\newcommand{\be}{\begin{equation}}
\newcommand{\ee}{\end{equation}}
\newcommand{\bea}{\begin{eqnarray}}
\newcommand{\eea}{\end{eqnarray}}

\begin{document}
\title{Semiclassical transport of hadrons with dynamical
spectral functions in $A+A$ collisions at SIS/AGS
energies\footnote{supported by GSI Darmstadt}}
\author{W. Cassing and S. Juchem\\
Institut f\"ur Theoretische Physik, Universit\"at Giessen\\
35392 Giessen, Germany}
\date{  }
\maketitle
\begin{abstract}
The transport of hadrons with dynamical spectral functions
$A_h(X,\vec{P},M^2)$ is studied for nucleus-nucleus collisions at
SIS and AGS energies in comparison to the conventional
quasi-particle limit and the available experimental data within
the recently developed off-shell HSD transport approach. Similar
to reactions at GANIL energies the off-shell effects show up in high
momentum tails of the particle spectra, however, at SIS and AGS energies
these modifications are found to be less pronounced than at lower energies
due to the high lying excitations of the nucleons in the collision zone.
\end{abstract}

\vspace{2cm}


\newpage
\section{Introduction}
\label{introduction} Nowadays, the dynamical description of
strongly interacting systems out of equilibrium is dominantly
based on transport theories and efficient numerical recipies have
been set up for the solution of the coupled channel transport
equations \cite{Stoecker,Bertsch,CMMN,Cass,T3,URQMD,CB99} (and
Refs. therein). These transport approaches have been derived
either from the Kadanoff-Baym equations \cite{kb62} in
Refs. \cite{pd841,Bot,Mal,ph95,gl98} or from the hierarchy of connected
equal-time Green functions \cite{Wang1,Zuo} in
Refs. \cite{Cass,CaWa,CNW} by applying a Wigner-transformation and
restricting to first order in the derivatives of the phase-space
variables ($X, P$).

However, as recognized early in these derivations \cite{Mal,CaWa},
the on-shell quasiparticle limit, that invoked additionally a
reduction of the $8N$-dimensional phase-space to $7N$ independent
degrees of freedom, where $N$ denotes the number of particles in
the system, should not be adequate for particles with high
collision rates (cf. also Refs. \cite{Knoll,Knoll2}). Therefore,
transport formulations for quasiparticles with dynamical spectral
functions have been presented in the past \cite{Bot,ph95}
providing a formal basis for an extension of the presently applied
transport models denoted by BUU/VUU
\cite{Bertsch,CMMN,Koreview,Niita,Wolf90}, QMD \cite{Aich,T1,T2}
or its relativistic versions RBUU \cite{Tomo}, UrQMD \cite{URQMD},
ART \cite{ART}, ARC \cite{Kah2} or HSD \cite{CB99,Ehehalt}.

Recently, the authors have developed a semiclassical transport
approach on the basis of the Kadanoff-Baym equations that includes
the propagation of hadrons with dynamical spectral functions
\cite{Cass99}. This approach has been examined for nucleus-nucleus
collisions at GANIL energies and the off-shell propagation of
nucleons and $\Delta$'s has lead to an enhancement of the high
energy proton spectra as well as to an enhancement of high energy
$\gamma$-rays rather well in line with related experimental
studies. The question now arises whether such phenomena will also
be encountered at SIS energies or AGS energies, i.e. at laboratory
energies that are higher by one or two orders of magnitude,
respectively.

The paper is organized as follows: In Section 2 we will briefly
review the generalized transport equations on the basis of the
Kadanoff-Baym equations \cite{kb62}, extend the test-particle
representation to the general case of momentum-dependent self energies and
specify the collision terms for bosons and fermions. Furthermore,
we will present the mass differential cross sections for mesons
from $NN$ and $\pi N$ collisions in the medium as well as
the approximations for their 'collisional broadening'. The results
for particle production in nucleus-nucleus collisions from the
off-shell approach will be discussed in Section 3 in comparison to
the quasi-particle limit as well as to the experimental data
available. A summary and discussion of open problems concludes
this study in Section 4.

\section{Extended semiclassical transport equations}
In this Section we briefly recall the basic equations for Green functions
and particle self energies as well as their symmetry properties that will be
exploited in the derivation of transport equations in the semiclassical limit.

The general starting point for the derivation of a transport equation for
particles with finite width are the Dyson-Schwinger equations for the
retarded and advanced Green functions $S^{ret}$, $S^{adv}$
and for the non-ordered Green functions $S^{<}$ and $S^{>}$ \cite{Cass99}.
In the case of scalar bosons -- which is considered in the following
for simplicity -- these Green functions are defined by \\
\bea
\hspace{2.0cm}
i \, S^{<}_{xy} & := & \; < \, \Phi^{\dagger}(y) \; \Phi(x) \, > \, ,
\nonumber\\
i \, S^{>}_{xy} & := & \; <\,  \Phi(x) \; \Phi^{\dagger}(y) \, > \, ,
\nonumber\\
i \, S^{ret}_{xy} & := & \phantom{-} \: \Theta (x_0 - y_0) \,
< \, [ \, \Phi(x) \, , \, \Phi^{\dagger}(y) \, ] \, > \, ,
\nonumber\\
i \, S^{adv}_{xy} & := & - \: \Theta (y_0 - x_0) \,
< \, [ \, \Phi(x) \, , \, \Phi^{\dagger}(y) \, ] \, > \, .
\eea\\
They depend on the space-time coordinates $x,y$ as indicated by the
indices $\cdot_{xy}$.
The Green functions are determined via Dyson-Schwinger equations by the
retarded and advanced self energies $\Sigma^{ret},\Sigma^{adv}$
and the collisional self energy $\Sigma^{<}$: \\
\bea
\hat{S}_{0x}^{-1} \; S_{xy}^{ret}
\; \: = \; \:
\delta_{xy}
\; \: + \; \:
\Sigma_{xz}^{ret} \: \odot \: S_{zy}^{ret} \; ,
\label{dsret_spatial}
\eea
\bea
\hat{S}_{0x}^{-1} \; S_{xy}^{adv}
\; \: = \; \:
\delta_{xy}
\; \: + \; \:
\Sigma_{xz}^{adv} \: \odot \: S_{zy}^{adv} \; ,
\label{dsadv_spatial}
\eea
\bea
\hat{S}_{0x}^{-1} \; S_{xy}^{<}
\; \: = \; \:
\Sigma_{xz}^{ret} \: \odot \: S_{zy}^{<}
\; \: + \; \:
\Sigma_{xz}^{<} \: \odot \: S_{zy}^{adv} \: .
\label{kb_spatial}
\eea \\
Equation (\ref{kb_spatial}) is the well-known Kadanoff-Baym equation.
Here $\hat{S}^{-1}_{0x}$ denotes the (negative) Klein-Gordon differential
operator which is given for bosonic field quanta of (bare) mass $M_0$ by
$\hat{S}^{-1}_{0x} = - (\partial^{\mu}_{x} \partial^{x}_{\mu} + M^2_0 )$,
$\delta_{xy}$ represents the four-dimensional $\delta$-distribution
$\delta_{xy} \equiv \delta^{(4)}(x-y)$ and the symbol $\odot$ indicates
an integration (from $-\infty$ to $\infty$) over all common intermediate
variables (cf. \cite{Cass99}).

For the derivation of a semiclassical transport equation one now
changes from a pure space-time formulation into the Wigner-representation.
The theory is then formulated in terms of the center-of-mass variable
$X = (x+y)/2$ and the momentum $P$, which is introduced by
Fourier-transformation with respect to the relative space-time coordinate
$(x-y)$.
In any semiclassical transport theory it is furthermore assumed that
the dependence on the mean space-time coordinates $X$ of all functions
is rather weak.
Therefore, in the Wigner-transformed expressions only contributions
up to the first order in the space-time gradients are kept.
After carrying-out these two steps the Dyson-Schwinger equations
(\ref{dsret_spatial}-\ref{kb_spatial}) become
\bea
\left[ \, P^2 \, - \, M^2_0 \, + \, i P^{\mu} \partial^{X}_{\mu} \, \right]
\: S^{ret}_{XP}
\; = \;
1 \: + \: ( \, 1 \, - \, i \, \Diamond \, ) \,
\{ \, \Sigma^{ret}_{XP} \, \} \: \{ \, S^{ret}_{XP} \, \} \: ,
\label{dsret_wignerfo}
\eea
\bea
\left[ \, P^2 \, - \, M^2_0 \, + \, i P^{\mu} \partial^{X}_{\mu} \, \right]
\: S^{adv}_{XP}
\; = \;
1 \: + \: ( \, 1 \, - \, i \, \Diamond \, ) \,
\{ \, \Sigma^{adv}_{XP} \, \} \: \{ \, S^{adv}_{XP} \, \} \: ,
\label{dsadv_wignerfo}
\eea
\bea
\left[ \, P^2 \, - \, M^2_0 \, + \, i P^{\mu} \partial^{X}_{\mu} \, \right]
\: S^{<}_{XP}
\; = \;
( \, 1 \, - \, i \, \Diamond \, ) \,
\left[ \: \{ \, \Sigma^{ret}_{XP} \, \} \: \{ \, S^{<}_{XP} \, \}
\; + \;
% ( \, 1 \, - \, i \, \Diamond \, ) \,
\{ \, \Sigma^{<}_{XP} \, \} \: \{ \, S^{adv}_{XP} \, \}
\: \right] .
\label{kb_wignerfo}
\eea\\
The operator $\Diamond$ is defined as \cite{ph95,Cass99}
\bea
 \Diamond \, \{ \, F_{1} \, \} \, \{ \, F_{2} \, \}
\; := \; \frac{1}{2} \left( \frac{\partial F_{1}}{\partial
X^{\mu}} \: \frac{\partial F_{2}}{\partial P_{\mu}} \; - \;
\frac{\partial F_{1}}{\partial P_{\mu}} \: \frac{\partial
F_{2}}{\partial X^{\mu}} \right) ,\label{poissonoperator} \eea\\
which is a four-dimensional generalization of the well-known
Poisson-bracket. Starting from (\ref{dsret_wignerfo}) and
(\ref{dsadv_wignerfo}) one obtains algebraic relations between the
real and the imaginary part of the retarded Green functions. On
the other hand eq. (\ref{kb_wignerfo}) leads to a 'transport
equation' for the Green function $S^{<}$.

We briefly recall the necessary steps: we separate all retarded
and advanced quantities -- Green functions and self energies  --
into real and imaginary parts,
\bea
S_{XP}^{ret,adv}
\; = \;
%G_{XP}
Re S^{ret}_{XP}
\; \mp \;
\frac{i}{2} \, A_{XP}
\; , \qquad
\Sigma_{XP}^{ret,adv}
\; = \;
Re \Sigma^{ret}_{XP}
\; \mp \;
\frac{i}{2} \, \Gamma_{XP} \; .
\label{ret_sep}
\eea\\
The imaginary part of the retarded propagator is given (up to a factor 2)
by the normalized spectral function
\bea A_{XP} \: = \: i \left[ \, S_{XP}^{ret} \: - \: S_{XP}^{adv}
\, \right] \: = \: - 2 \, Im \, S^{ret}_{XP} \; , \qquad \qquad
\qquad \int \frac{d P_0^2}{4 \pi} \,  A_{XP} \; = \; 1 \; ,
\label{spectralfunction} \eea\\
while the imaginary part of the self energy corresponds to half the
width $\Gamma_{XP}$.
By separating the complex equations (\ref{dsret_wignerfo}) and
(\ref{dsadv_wignerfo}) into their real and imaginary contributions
we obtain an algebraic equation between the real and the imaginary
part of $S^{ret}$,
\bea
Re S^{ret}_{XP} \; = \;
\frac{P^{2} \: - \: M_{0}^{2} \: - \: Re \Sigma^{ret}_{XP}}{\Gamma_{XP}}
\; A_{XP} \, .
\label{dispersion}
\eea\\
In addition we gain an algebraic solution for the spectral function
(in first order gradient expansion) as
\bea A_{XP} \; = \; \frac{ \Gamma_{XP} } {( \, P^2 \, - \,
M_{0}^{2} \, - \, Re \Sigma^{ret}_{XP} )^{2} \: + \:
\Gamma_{XP}^{2}/4} \; , \label{alg_spectral} \eea\\
while the real part of the retarded propagator is given by
\bea Re S^{ret}_{XP} \; = \; \frac{P^{2} \: - \: M_{0}^{2} \: - \:
Re \Sigma^{ret}_{XP}} {( \, P^2 \, - \, M_{0}^{2} \, - \, Re
\Sigma^{ret}_{XP} )^{2} \: + \: \Gamma_{XP}^{2}/4} \, .
\label{alg_realpart} \eea \\
Furthermore, the (Wigner-transformed) Kadanoff-Baym equation
(\ref{kb_wignerfo}) allows for the construction of a transport
equation for the Green function $S^{<}$.
When separating the real and the imaginary contribution of this
equation we find i) a generalized transport equation,
\bea
\Diamond \, \{ \, P^{2} &-& M_{0}^{2} \: - \: Re \Sigma^{ret}_{XP} \, \}
\; \{ \, S^{<}_{XP} \, \}
\; - \;
\Diamond \, \{ \, \Sigma^{<}_{XP} \, \} \; \{ Re S^{ret}_{XP} \, \}
\nonumber\\[0.2cm]
&=&
\frac{i}{2} \:
\left[ \: \Sigma^{>}_{XP} \: S^{<}_{XP} \; - \;
          \Sigma^{<}_{XP} \: S^{>}_{XP} \: \right],
\label{general_transport}
\eea\\
and ii) a generalized mass-shell constraint
\bea
[ \, P^{2} &-& M_{0}^{2} \: - \: Re \Sigma^{ret}_{XP} \, ] \; S^{<}_{XP}
\; - \;
\Sigma^{<}_{XP} \; Re S^{ret}_{XP}
\nonumber\\[0.2cm]
&=&
\frac{1}{2} \,
\Diamond \; \{ \, \Sigma^{<}_{XP} \, \} \; \{ \, A_{XP} \, \}
\; - \;
\frac{1}{2} \,
\Diamond \; \{ \, \Gamma_{XP} \, \} \; \{ \, S^{<}_{XP} \, \} \; .
\label{general_massshell}
\eea\\
We note that so far no reduction to the quasiparticle limit has
been introduced. Besides the drift term (i.e. $\Diamond \{P^2 -
M^2_0\} \{ S^{<} \} = - P^{\mu} \partial^{X}_{\mu} S^{<} )$ and the
Vlasov term (i.e. $\Diamond \{ Re \Sigma^{ret} \} \{ S^{<} \} $) a
third contribution appears on the l.h.s. of
(\ref{general_transport}) (i.e. $\Diamond \{\Sigma^{<} \} \{Re
S^{ret} \}$), which vanishes in the quasiparticle limit and
incorporates -- as shown in \cite{Cass99} -- the off-shell
behaviour in the particle propagation. It is worth to point out
that this particular term has been discarded in Ref. \cite{Effe}
and an off-shell propagation had to be introduced by hand via
some mass-dependent real potential. To our best knowledge, such an
auxiliary potential cannot be extracted from the Kadanoff-Baym
equation (\ref{kb_spatial}). The r.h.s. of
(\ref{general_transport}) consists of a collision term with its
characteristic gain ($\sim \Sigma^{<} S^{>}$) and loss ($\sim
\Sigma^{>} S^{<}$) structure, where scattering processes of
particles into and out off a given phase-space cell are described.

Within the specific term ($\Diamond \{\Sigma^{<} \} \{Re S^{ret}
\}$) a further modification is necessary. According to Botermans
and Malfliet \cite{Bot} the collisional self energy $\Sigma^{<}$
should be replaced by $S^{<} \cdot \Gamma / A$ to gain a
consistent first order gradient expansion scheme. The replacement
is allowed since the difference between these two expressions can
be shown to be of first order in the space-time gradients itself.
Therefore it has to be discarded when appearing inside an
additional Poisson-bracket. Furthermore, this substitution is
required to get rid of the inequivalence between the general
transport equation and the general mass shell constraint which is
present in (\ref{general_transport}) and (\ref{general_massshell})
\cite{Knoll} but vanishes by invoking $S^{<} \cdot \Gamma / A$.
Finally, the general transport equation (in first order gradient
expansion) reads \cite{Cass99}
\setlength{\mathindent}{-0.5cm}
\bea A_{XP} \, \Gamma_{XP} && \!\!\!\!\! \!\!\!\!\! \left[ \,
\Diamond \; \{ \, P^2 - M_0^2 - Re \Sigma^{ret}_{XP} \, \} \; \{
\, S^<_{XP} \, \} \: - \: \frac{1}{\Gamma_{XP}} \; \Diamond
\; \{ \, \Gamma_{XP} \, \} \; \{ \, ( \, P^2 - M_0^2 - Re
\Sigma^{ret}_{XP} \, ) \, S^<_{XP} \, \} \, \right]
\nonumber\\[0.2cm] && \; = \; i \, \left[ \, \Sigma^>_{XP} \:
S^<_{XP} \: - \: \Sigma^<_{XP} \: S^>_{XP} \, \right].
\label{trans_approx} \eea \\
\setlength{\mathindent}{0.5cm}
Its formal structure is fixed by the approximations applied. Note,
however, that the dynamics is fully determined by the different
self energies, i.e. $Re \Sigma_{XP}^{ret}, \Gamma_{XP},
\Sigma_{XP}^<$ and $\Sigma_{XP}^>$ that have to be specified for
the physical systems of interest.

\subsection{Testparticle representation}

In order to obtain an approximate solution to the transport
equation (\ref{trans_approx}) we use a testparticle ansatz for the
Green function $S^{<}$, more specifically for the real and
positive semidefinite quantity
\bea
F_{XP} \; = A_{XP} N_{XP} = \; i \, S^{<}_{XP}
\; \sim \; \sum_{i=1}^{N} \; \delta^{(3)} ({\vec{X}} \, - \,
{\vec{X}}_i (t)) \; \; \delta^{(3)} ({\vec{P}} \, - \, {\vec{P}}_i
(t)) \; \; \delta(P_0 - \, \epsilon_i(t)) \: .
\label{testparticle} \eea \\
Whereas so far we have briefly repeated the derivation from
\cite{Cass99}, we now extend the testparticle description to
explicitly four-momentum-dependent self energies. The latter is
not essential for the description of low and intermediate energy heavy-ion
reactions, but becomes necessary for relativistic dynamics. In the
most general case (where the self energies depend on four-momentum
$P$, time $t$ and the spatial coordinates $\vec{X}$) the equations of motion
for the testparticles  read
\bea
\label{eomr}
\frac{d {\vec X}_i}{dt} \! & = & \!
\phantom{- }
\frac{1}{1 - C_{(i)}} \,
\frac{1}{2 \epsilon_i} \:
\left[ \, 2 \, {\vec P}_i \, + \, {\vec \nabla}_{P_i} \, Re \Sigma^{ret}_{(i)}
\, + \, \frac{ \epsilon_i^2 - {\vec P}_i^2 - M_0^2
- Re \Sigma^{ret}_{(i)}}{\Gamma_{(i)}}
\: {\vec \nabla}_{P_i} \, \Gamma_{(i)} \:
\right],
\\[0.3cm]
\label{eomp}
\frac{d {\vec P}_i}{d t} \! & = & \!
- \frac{1}{1-C_{(i)}} \,
\frac{1}{2 \epsilon_{i}} \:
\left[ {\vec \nabla}_{X_i} \, Re \Sigma^{ret}_i
\: + \: \frac{\epsilon_i^2 - {\vec P}_i^2 - M_0^{2}
- Re \Sigma^{ret}_{(i)}}{\Gamma_{(i)}}
\: {\vec \nabla}_{X_i} \, \Gamma_{(i)} \:
\right],
\\[0.3cm]
\label{eome}
\frac{d \epsilon_i}{d t} \! \setlength{\mathindent}{-0.5cm}
& = & \!
\phantom{- }
\frac{1}{1 - C_{(i)}} \,
\frac{1}{2 \epsilon_i} \:
\left[ \frac{\partial Re \Sigma^{ret}_{(i)}}{\partial t}
\: + \: \frac{\epsilon_i^2 - {\vec P}_i^2 - M_0^{2}
- Re \Sigma^{ret}_{(i)}}{\Gamma_{(i)}}
\: \frac{\partial \Gamma_{(i)}}{\partial t}
\right],
\eea\\
where the notation $F_{(i)}$ implies that the function is taken at
the coordinates of the testparticle, i.e.
$F_{(i)} \equiv F(t,\vec{X}_{i}(t),\vec{P}_{i}(t),\epsilon_{i}(t))$.

In (\ref{eomr}-\ref{eome}) a common multiplication factor $(1-C_{(i)})^{-1}$
appears, which contains the energy derivatives of the retarded
self energy\\
\bea
\label{correc}
C_{(i)} \: = \:
\frac{1}{2 \epsilon_i}
\left[
\frac{\partial}{\partial \epsilon_i} \, Re \Sigma^{ret}_{(i)} \: + \:
\frac{\epsilon_i^2 - {\vec P}_i^2 - M_0^2
- Re \Sigma^{ret}_{(i)}}{\Gamma_{(i)}}
\: \frac{\partial }{\partial \epsilon_i} \,
\Gamma_{(i)}
\right] \: .
\eea\\
It yields a shift of the system time $t$ to the 'eigentime' of
particle $i$ defined by $\tilde{t}_{i} = t /(1-C_{(i)})$. As the
reader immediately verifies, the derivatives with respect to the
'eigentime', i.e. $d \vec{X}_i / d \tilde{t}_i$, $d \vec{P}_i / d
\tilde{t}_i$ and $d \epsilon_i / d \tilde{t}_i$ then emerge
without this renormalization factor for each testparticle
$i$ when neglecting higher order time derivatives in line with the
semiclassical approximation scheme.
The role and the importance of this correction factors will
be studied in detail for a four-momentum-dependent 'trial'
potential in Section \ref{trialpot}. We note that for
momentum-independent self energies we regain the transport
equations as derived in \cite{Cass99}; only in case of particles
with a vanishing vacuum width $\Gamma_V$ = 0 and $\Gamma_{XP} \sim \rho_B$
(baryon density) these equations reduce to the 'ad hoc' assumptions
introduced in Ref. \cite{Effe}. Furthermore, in the limiting case of
particles with vanishing gradients of the width $\Gamma_{XP}$ these
equations of motion  reduce to the well-known transport equations
of the quasiparticle picture.

Following Ref. \cite{Cass99} we take $M^{2} = P^2 - Re \Sigma^{ret}$ as
an independent variable instead of $P_0$, which then fixes the energy
(for given $\vec{P}$ and $M^{2}$) to
\bea
P_{0}^{2} \; = \; \vec{P}^{2} \: + \: M^{2} \: + \:
Re \Sigma_{X\vec{P}M^2}^{ret} \, .
\label{energyfix}
\eea
Eq. (\ref{eome}) then turns to
\bea
\label{eomm}
\frac{dM_i^2}{dt} \; = \;
\frac{M_i^2 - M_0^2}{\Gamma_{(i)}} \;
\frac{d \Gamma_{(i)}}{dt}
\eea
for the time evolution of the test-particle $i$ in
the invariant mass squared as derived in Ref. \cite{Cass99}.

We briefly comment that in Ref. \cite{Cass99} we have added a term
$\sim \partial \Gamma_{(i)}/\partial t$ in the equation of motion
for the test-particle momenta in order to achieve strict energy
conservation for each test-particle; however, within the
subsequent approximations introduced in the actual transport
calculations for energetic nucleus-nucleus collisions such a term
did not lead to noticable effects within the numerical accuracy
achieved. We thus discard such a term in the present formulation and
investigation.

\subsection{Model simulations for the momentum-dependent transport
equations in testparticle representation} \label{trialpot} To
demonstrate the physical content of the equations of motion for
testparticles (\ref{eomr}-\ref{eome}) we perform an exploratory
study with a momentum-dependent trial potential. The potential is
chosen of the type:
\bea
Re \Sigma^{ret} - \frac{i}{2} \Gamma
\: = \:
\frac{V(P_0,\vec{P})}{1+\exp\{(|\vec{r}|-R)/a_0\}}
\; - \; i \left(
\frac{W(P_0,\vec{P})}{1+\exp\{(|\vec{r}|-R)/a_0\}}
\: + \: \frac{\Gamma_V}{2} \right)
\eea
with a constant (but finite) vacuum width $\Gamma_V$. While the
spatial extension of the potential is given (as in \cite{Cass99})
by a Woods-Saxon shape (with parameters $R = 5$ fm and $a_0 = 0.6$ fm)
its momentum dependence for the real as well as for the imaginary
part is introduced by
\bea
V(P_0,\vec{P}) \; \; = C_V \:
\frac{\Lambda_V^2}{\Lambda_V^2 - (P_0^2 - \vec{P}^2)} \: , \qquad
W(P_0,\vec{P}) \; \; = C_W \:
\frac{\Lambda_W^2}{\Lambda_W^2 - (P_0^2 - \vec{P}^2)}
\label{momdeppart} \: .
\eea
Here the constants $C_V$ ($C_W$) give the 'strength' of the
complex potential while $\Lambda_V$ ($\Lambda_W$) play the role of
cutoff-parameters. Due to the structure in the denominator of
(\ref{momdeppart}) the momentum-dependent part of this potential
is explicitly Lorentz-covariant.

In our simulation we propagate the testparticles with different
initial mass parameters $M_i$, which are shifted relative to each
other by $\Gamma_V/(20$ GeV) around a mean mass of 1.0 GeV. To
each testparticle a momentum in positive $z$-direction is
attributed so that all of them have initially the same energy
$P_0$ = 2.0 GeV. All testparticles are initialized on the negative
$z$-axis with ($|\vec{X}_i(t=0)| \approx 15$ fm) and then evolved
in time according to the equations of motion
(\ref{eomr}-\ref{eome}).

In our first simulation we consider a purely imaginary potential
with a strength of $C_W = 0.6$ GeV$^2$ and a cutoff-parameter
$\Lambda_W = 2.0$ GeV. The evolution in energy $P_{0i}$, momentum
$P_{zi}$ and in the mass parameter $M_i$ for all testparticles is
shown in Fig. 1 (upper part) as a function of $z(t)$. When the
testparticles enter the potential region, their momenta and mass
parameters are modified. As already shown in \cite{Cass99} the
imaginary potential leads to a spreading of the trajectories in
the mass parameter $M_i$ which in turn reflects a broadening of
the spectral function. The relation between the imaginary self
energy and the spreading in mass is fully determined by relation
(\ref{eomm}). Since we have chosen a potential with no explicit
time dependence the energy of each testparticle is a constant of
time (cf. Ref. \cite{Cass99}). According to the explicit momentum
dependence of our 'trial' potential each single testparticle is
affected with different strength. Since the imaginary potential is
strongest for small momenta (which correspond to the highest lines
in the lower graph of Fig. 1) the momentum and mass coordinates of
those testparticles are changed predominantly that are initialized
with the lowest momenta (i.e. with the largest masses). As a
result one observes a rather asymmetric distribution in the mass
parameters (and in the momenta) in the potential zone. This is
different from the studies of \cite{Cass99} where the
investigated momentum-independent potential yields a nearly
equidistant spreading of the mass trajectories. For $z(t) \gg R$ the
mass and momentum coordinates of the testparticles return to the
proper asymptotic value.

In the second example we allow for an addititonal real part of the
self energy. The calculation is performed with the parameters $C_V
= -0.3$ GeV$^2$, $C_W = 0.6$ GeV$^2$ and $\Lambda_V = \Lambda_W =
2.0$ GeV. The momentum-dependent real part (lower part of Fig. 2)
causes -- as also observed in \cite{Cass99} -- an additional shift
of the testparticle momenta. Since the real part of the potential
is larger for small initial momenta, these testparticle momenta
are shifted up somewhat more than for particles with larger
momenta. This gives rise to a reduction of the asymmetry which was
introduced by the momentum-dependent imaginary part of the self
energy (upper part of Fig. 2). As in \cite{Cass99} the mass parameters of the
testparticles are only weakly influenced by the real part of the
potential.

We, furthermore, study the implications of the correction term
$(1-C_{(i)})^{-1}$ using the same imaginary potential as in Fig. 1. The
time evolution of the correction factor for each testparticle $i$
is displayed in the lower part of Fig 3. While it is $> 1$ for
large initial mass parameters $M_i > M_0$, it is $ < 1$ for mass
parameters $M_i < M_0$. In the upper part of Fig. 3 the momenta of
the testparticles are shown for two calculational limits: in the
first one the correction term is taken into account (as in the
previous calculations), while in the second one the corrections
due to the energy dependence of the retarded self energy are
neglected. However, the calculations with and without the
correction factor exhibit only a very small difference in the
testparticle momenta, which even cannot be distinguished within
the resolution of Fig. 3. The same holds for the mass parameters
$M_i$ which are not displayed here since they provide no new information
due to energy conservation. We thus conclude that the particle
trajectory is not very sensitive to these correction factors,
since the correction term -- when displayed as $P_{z}(z)$ in
phase-space -- leads only to a rescaling of the 'eigentime' of the
testparticles as pointed out before.

\subsection{Collision terms}
The collision term of the Kadanoff-Baym equation can only be
worked out in more detail by giving explicit approximations for
$\Sigma^{<}$ and $\Sigma^{>}$. A corresponding collision term can
be formulated in full analogy to Refs. \cite{CMMN,CaWa}, e.g. from
Dirac-Brueckner theory, and implementing detailed balance as
$$
I_{coll}(X,\vec{P},M^2) = Tr_2 Tr_3 Tr_4 A(X,{\vec P},M^2)
A(X,{\vec P}_2, M_2 ^2) A(X,{\vec P}_3, M_3 ^2)
A(X,{\vec P}_4, M_4 ^2)
$$
$$
%\{
|T(({\vec P},M^2) + ({\vec P}_2,M_2^2)
\rightarrow ({\vec P}_3,M_3^2) + ({\vec P}_4,M_4^2))|_{{\cal A,S}}^2
\; \; \delta^{(4)}({P} + {P}_2 - {P}_3 - {P}_4)
$$
\be
\label{Icoll}
[ \, N_{X{\vec P}_3 M_3^2} \, N_{X {\vec P}_4 M_4^2} \,
{\bar f}_{X {\vec P} M^2} \, {\bar f}_{X {\vec P}_2 M_2^2}
\: - \: N_{X{\vec P} M^2} \, N_{X {\vec P}_2 M_2^2} \,
{\bar f}_{X {\vec P}_3 M_3^2} \,
{\bar f}_{X {\vec P}_4 M_4^2} \, ]
%\}
\ee
with
\be
\label{pauli} {\bar f}_{X {\vec P} M^2} = 1 + \eta \, N_{X {\vec P} M^2} \ee
and $\eta = \pm 1$ for bosons/fermions, respectively. The indices
${\cal A,S}$ stand for the antisymmetric/symmetric matrix element
of the in-medium scattering amplitude $T$ in case of
fermions/bosons. In eq. (\ref{Icoll}) the trace over particles
2,3,4 reads explicitly for fermions
\be
\label{trace}
Tr_2 = \sum_{\sigma_2, \tau_2} \frac{1}{(2 \pi)^4}
\int d^3 P_2 \frac{d M^2_2}{2 \sqrt{\vec{P}^2_2+M^2_2}},
\ee
where $\sigma_2, \tau_2$ denote the spin and isospin of particle 2.
In case of bosons we have
\be
\label{trace2} Tr_2 = \sum_{\sigma_2, \tau_2} \frac{1}{(2 \pi)^4}
\int d^3 P_2 \frac{d P_{0,2}^2}{2}, \ee since here the spectral
function $A_B$ is normalized as
\be
\label{sb} \int \frac{d P_0^2}{4 \pi} A_B(X,P) = 1 \ee whereas for
fermions we have
\be
\label{sb1}
\int \frac{d P_0}{2 \pi} A_F(X,P) = 1.
\ee
We mention that the spectral function $A_F$ in case of fermions in
(\ref{Icoll}) is obtained
by considering only particles of positive energy and assuming the spectral
function to be identical for spin 'up' and 'down' states. In general, the
spectral function for fermions ${\hat A}_{\alpha \beta}(X,P)$ is a Dirac-tensor
with $\alpha \beta$ denoting the Dirac indices. It is normalized as
\be
\label{specF}
\int \frac{d P_0}{2 \pi} {\hat A}_{\alpha \beta} (X,P) =
(\gamma^0)_{\alpha \beta},
\ee
which implies
\be
\frac{1}{4} \sum_{\alpha} \int \frac{d P_0}{2 \pi} (\gamma^0 {\hat
A}(X,P))_{\alpha \alpha} = 1. \ee Now expanding ${\hat A}$ in
terms of free spinors $u_s(P)$ ($s$=1,2) and $v_s(P)$ as
\be
(\hat{A})_{\alpha \beta} \: = \:
\sum_{r,s=1}^{2}
\bar{u}_s(\vec{P}_,M)_{\beta} \; A_{rs}^{p} \; u_r(\vec{P},M)_{\alpha}
\: + \:
\bar{v}_s(\vec{P}_,M)_{\beta} \; A_{rs}^{ap} \; v_r(\vec{P},M)_{\alpha}
\ee
one can separate particles and antiparticles. By neglection of the
antiparticle contributions (i.e. $A_{rs}^{ap} \equiv 0$) and within the
assumption that the spectral function for the particles is diagonal
in spin-space (i.e. $A_{rs}^{p} = \delta_{rs} A_s^{p}$) as well as
spin symmetric, one can define $A_F$ as
\be
A_F \equiv A_1^{p} = A_2^{p} \: .
\ee

Neglecting the 'gain-term' in eq. (\ref{Icoll}) one recognizes
that the collisional width of the particle in the rest frame is given by
\be
\label{gcoll}
\Gamma_{coll}(X,\vec{P},M^2) = Tr_2 Tr_3 Tr_4 \;
%\{
|T(({\vec P},M^2) + ({\vec P}_2,M_2^2) \rightarrow ({\vec
P}_3,M_3^2) + ({\vec P}_4,M_4^2))|_{{\cal A,S}}^2 \ee $$ A(X,{\vec
P}_2,M_2^2) A(X,{\vec P}_3,M_3^2) A(X,{\vec P}_4, M_4^2)
\; \; \delta^4(P + P_2 - P_3-P_4) \ N_{X {\vec P}_2 M_2^2} \,
{\bar f}_{X {\vec P}_3 M^2_3} \, {\bar f}_{X {\vec P}_4 M^2_4} \,
%\}
, $$ where as in
eq. (\ref{Icoll}) local on-shell scattering processes are assumed for
the transitions $P + P_2 \rightarrow P_3 + P_4$. We note that the
extension of eq. (\ref{Icoll}) to inelastic scattering processes
(e.g. $NN \rightarrow N\Delta$) or ($\pi N \rightarrow \Delta$
etc.) is straightforward when exchanging the elastic transition
amplitude $T$ by the corresponding inelastic one and taking care
of Pauli-blocking or Bose-enhancement for the particles in the
final state. We note that for bosons we will neglect a
Bose-enhancement factor throughout this work since their actual
phase-space density is small for the systems of interest.

For particles of infinite life time in vacuum -- such as protons
-- the collisional width (\ref{gcoll}) has to be identified with
twice the imaginary part of the self energy. Thus the transport
approach determines the particle spectral function dynamically via
(\ref{gcoll}) for all hadrons if the in-medium transition
amplitudes $T$ are known {\it in their full off-shell dependence}.
Since this information is not available for configurations of hot
and dense matter, which is the major subject of future
development, a couple of assumptions and numerical approximation
schemes have to be invoked in actual applications.

\subsection{Numerical realization}
As in Ref. \cite{Cass99} the
following dynamical calculations are based on the conventional HSD
transport approach \cite{CB99,Ehehalt} -- in which $Re
\Sigma^{ret}_{XP}$ is specified for the hadrons -- however, the
equations of motion for the testparticles are extended to
(\ref{eomr},\ref{eomp},\ref{eomm}). Whereas for energies up to 100
A MeV (GANIL energies) essentially the nucleon degrees of freedom
were important \cite{Cass99} we have to specify the actual
'recipies' for nucleon and meson resonances involved in the
following calculations.

As a first approximation we only consider reactions with binary
channels which for convenience can be described in their
center-of-mass (cms) frame. The collisions of nucleons (as well as
all hadrons) are described by the closest distance criterion of
Kodama et al. \cite{Kodama}: a collision of two particles takes
place only if their distance in the individual cms is small
enough, i.e.
\be
|{\vec X}_1 - {\vec X}_2| \leq \sqrt{\sigma({|\vec P}_1 - {\vec
P}_2|)/\pi}, \label{collcrit} \ee
where $\sigma$ denotes the total cross section of the process,
which is written here as a function of the three-momentum
difference in the cms. In the case of nucleon-nucleon collisions
the Cugnon parametrization \cite{Cugnon} for the in-medium $NN$
cross section $d\sigma/d\Omega(\sqrt{s'})$ is used by identifying
(in the $NN$ c.m.s.)
\be
s'- 4 m_N^2 = s-4M^2 = 4 {\vec P}^2, \ee where $m_N$ is the
nucleon vacuum mass, $M$ the actual off-shell mass and $\sqrt{s'}$
the invariant energy of a nucleon-nucleon collision in the vacuum
with the same cms-momentum ${\vec P}$. The final nucleon states
are selected by Monte-Carlo according to the local spectral
function determined by the collisional width $\Gamma_{coll}(X,P)$,
while the angular distribution in the cms is taken the same as for
on-shell nucleons. This recipe for off-shell nucleon-nucleon
scattering is practically an {\it ad-hoc} assumption and has to be
controlled by off-shell matrix elements of the nucleon Brueckner
T-matrix in the medium. In Section 3 we will also investigate an
alternative recipe to demonstrate the model dependence of our
actual results.

If two particles have approached sufficiently close, furthermore,
the time of the collision in the calculational frame has to be
specified. Within a relativistic treatment this is a nontrivial
task since 'simultaneity' is a non-invariant property. Therefore,
it is necessary to take into account the time coordinates $t_1$
and $t_2$ for both participants of the collision separately. By
determining the collision time (as the time of the closest
approach) in the individual cms of each colliding particle, a
causality respecting time ordering of all collisions in the
calculational frame can be achieved. Without representing the
explicit formulae  we refer the reader for a detailed description
to eqs. (8)-(11) of Kodama et al. \cite{Kodama} which are
implemented in the HSD transport approach.

In order to determine the cross section $\sigma$ in
(\ref{collcrit}) let us consider a binary reaction of hadrons
characterized by ${\vec P}_1, M^2_1$ and ${\vec P}_2, M_2^2$
suppressing all internal quantum numbers. In case of a single
final state, i.e. a resonance $R$ for meson-nucleon or meson-meson
scattering, energy and momentum conservation fixes the
four-momentum of the final state completely. The cross section for
the production of the resonance $R$ in the center-of-mass system
with invariant energy $M$ then is given by the Breit-Wigner cross
section
\be
\label{BWC}
 \sigma(M^2) \: = \: C_{S,I} \: \frac{1}{p_{in}^2} \:
\frac{M^2 \Gamma_{tot}^2 B_{in}}{(M^2 - M_0^2)^2 + M^2
\Gamma_{tot}^2} ,\ee where $p^2_{in}$ is the squared
three-momentum of a particle in the entrance channel, $B_{in}$ the
branching ratio to the resonance $R$ and
\be
\label{widthm} \Gamma_{tot} = \Gamma_V(M^2) + \Gamma_{coll}(M^2)
\equiv \frac{\Gamma_{XP}}{2 P_0} \, .
\ee
In (\ref{widthm}) $\Gamma_V$ and $\Gamma_{coll}$ denote the
vacuum and collisional particle width, respectively;
it is related to $\Gamma_{XP} = -2 \, Im \Sigma^{ret}$ by
$\Gamma_{XP} = 2 P_0 \, \Gamma_{tot}$ \cite{Cass99}.
In (\ref{BWC}) the factor $C_{S,I}$ is the usual spin/isospin factor
determined by the spin/isopin in the entrance channel and the
resonance properties. The vacuum decay width for the $\Delta$ is
taken in the parametrization of \cite{Wolf90} corrected by the
ratio of the two-body phase-space integrals
$R_2(\sqrt{s},M_1,M_2)/R_2(\sqrt{s},M_1^0,M_2^0)$.
Here the two-body phase-space integral is given by \cite{Byc73}
\bea
R_2(s,m^2_1,m^2_2) \: = \: \frac{\pi}{2s} \lambda^{\frac{1}{2}}(s,m^2_1,m^2_2)
\label{phase_r2}
\eea
with
\bea
\lambda(s,m^2_1,m^2_2) \: = \:
\left[ \, s \, - \, ( m_1 + m_2 )^2 \, \right] \;
\left[ \, s \, - \, ( m_1 - m_2 )^2 \, \right] \: .
\label{phase_l}
\eea
For the $N(1440)$ we assume the same parametrization of the width
as in Ref. \cite{Wolf90} corrected again by the ratio of two-body
phase-space integrals. The same strategy is taken for the
$N(1535)$ \cite{Wolf90} which is important for $\eta$ production,
rescattering and absorption. Higher baryon resonances (as in Refs.
\cite{Teis,Effe99}) are not considered in this study since they
are not seen experimentally in the photon absorption experiments
in Frascati even on light nuclei \cite{Frascati}.

In case of binary exit channels the final masses $M_3, M_4$ are
selected by Monte-Carlo according to the local spectral functions
with width $\Gamma_{tot}$. The $NN \rightarrow NR$ transitions are
simulated by i) correcting the vacuum width according to the
available phase-space in the final channel and ii) by adding the
collisional width $\Gamma_{coll}$. These assumptions presently are
also hard to control by fully microscopic calculations and serve
as a guide for the effects to be investigated below.

In case of meson production by off-shell baryon-baryon or meson-baryon
collisions we either have 2 (e.g. $\pi N \rightarrow K^+ \Lambda/ \Sigma$),
3 (e.g. $NN \rightarrow K^+ \Lambda N$ or $K^+ \Sigma N$) or 4 particles
(e.g. $NN \rightarrow NN K^+ K^-$)
in the final channel. Since the final mesons may be off-shell as well, one has
to specify the corresponding mass-differential cross sections that depend on
the entrance channel and especially on the availabe energy $\sqrt{s}$ in the
entrance channel.

We start with the explicit parametrizations for meson ($m$)
production cross sections given in Ref. \cite{CB99} for on-shell
mesons as a function of the invariant energy $\sqrt{s}$ in case of
nucleon-nucleon or pion-nucleon collisions, i.e. $\sigma_{NN
\rightarrow NNm}(\sqrt{s})$ or $\sigma_{\pi N \rightarrow
mN}(\sqrt{s})$, respectively, that are well controlled by
experimental data. Far above the corresponding thresholds the mass
differential cross sections are approximated by
\be
\label{c1} \frac{d \sigma_{NN \rightarrow mNN}(\sqrt{s})}{d M^2} \: = \:
\sigma_{NN \rightarrow mNN}(\sqrt{s^{\phantom{*}}_{\phantom{0}}}
-\sqrt{s^{*}_0} \: ) \; A_m(M^2,
\Gamma_{tot}) , \ee
where $A_m(M^2, \Gamma_{tot})$ denotes the
meson spectral function for given total width $\Gamma_{tot}$ that
is normalized to unity by integration over $d M^2$. In (\ref{c1})
the threshold energy $\sqrt{s^*_0} = M_0 + M_1^*+M_2^*$ depends on
the masses of the hadrons in the final channel, i.e. $M_0, M_1^*$
and $M_2^*$. Actual events then are selected by Monte-Carlo
according to (\ref{c1}). Close to threshold $\sqrt{s^*_0}$, i.e. for
$\sqrt{s} - M_0 - M^*_1 - M^*_2 \leq 2 \Gamma_{tot}$, where
$M^*_1, M^*_2$  denote the final off-shell masses of two nucleons,
$M_0$ the meson pole mass and $\Gamma_{tot}$ its total width, the
differential production cross section is approximated by a
constant matrix element squared $|M_m|^2$ times available
phase-space,
\be
\label{c2}
\frac{d \sigma_{NN \rightarrow mNN}(\sqrt{s})}{d M^2} \: = \:
|M_m|^2 \; A_m(M^2, \Gamma_{tot}) \; R_3(s,M^2,M_1^{2*},M^{2*}_2).
\ee
The matrix element $|M_m|$ then is fitted to the on-shell cross section
close to threshold. In (\ref{c2}) the function $R_3$ denotes the 3-body
phase-space integral in case of a $mNN$ final state and is given by
\cite{Byc73}
\bea
R_3(s,m^2_1,m^2_2,m^2_3) \: = \:
\int^{(\sqrt{s}-m_1)^2}_{(m_2+m_3)^2}
\frac{ds_2}{s_2}
\; \lambda^{\frac{1}{2}}(s_2,s,m^2_1)
\; \lambda^{\frac{1}{2}}(s_2,m^2_2,m^2_3) \: .
\label{phase_r3}
\eea
The same recipe is used for binary $mN$ channels by replacing $R_3$
with the 2-body phase-space integral $R_2$.

In case of 4 particles in the final state, e.g. in the channel $NN
\rightarrow K \bar{K} N^*_1 N^*_2$, where the $N^*$'s denote
off-shell nucleons, the differential cross section is approximated
by  \bea E_1 E_2 E_3 E_4 \frac{d^{12} \sigma_{BB \rightarrow
NNM_1M_2}(\sqrt{s})}{ d^3p_1 d^3p_2 d^3p_3 d^3p_4} =
\\  \nonumber \sigma_{BB \rightarrow NN M_1M_2}(\sqrt{s})
\frac{1}{16 R_4(\sqrt{s})} \delta^4(P_1 + P_2 -p_1-p_2-p_3-p_4),
 \eea
where $R_4$ denotes the 4-body phase-space integral \cite{Byc73}. Similar
strategies have been exploited in case of subthreshold $p \bar{p}$
production in proton-nucleus and nucleus-nucleus collisions in
Refs. \cite{Teis1,Sib98}.

The resulting cross sections for $K^-$  production from $NN$ and
$\pi^+ p$ collisions are displayed in Fig. 4 as a function of
$\sqrt{s}$ for different collisional width $\Gamma_{coll}$= 0, 50,
100, 150, 200 MeV while keeping $\sqrt{s_0} = M_K + M_N +
M_{\Lambda}$ or $\sqrt{s_0} = M_K + M_{\Lambda}$, respectively.
With increasing width $\Gamma_{coll}$ the subthreshold production
of mesons becomes enhanced considerably relative to the respective
vacuum cross section, but the absolute magnitude stays small below
threshold even for $\Gamma_{coll}$ = 200 MeV.

As a next step we have to fix the collisional width of the hadrons
in the nuclear medium which enters the spectral function
$A(X,{\vec P},M^2)$ as well as the differential cross sections
(\ref{c1}). According to (\ref{gcoll}) the collisional width is
explicitly momentum- (and energy-) dependent. Whereas in Ref.
\cite{Cass99} we have employed momentum-independent collisional
broadening, this is no longer adequate for relativistic systems.
We thus evaluate $\Gamma_{coll}$ for each particle in a finite
cell in coordinate space according to (\ref{gcoll}), however,
discard the explicit dependence on the energy $P_0$. This
approximation implies that the correction factors
$(1-C_{(i)})^{-1}$ in the testparticle equations of motion (18) -- (20) are
equal to 1 and all particles can be propagated with the same
system time.

In case of kaons, antikaons or $\rho$ mesons at SIS energies we
treat the latter perturbatively as in Ref. \cite{Cass97}, i.e.
each testparticle achieves a weight $W_i$ defined by the ratio of
the individual production cross section to the total $\pi B$ or
$BB$ cross section at the same invariant energy. Their propagation
and interactions are evaluated as for baryons and pions, however,
the baryons (pions) are not changed in their final state when
interacting with a 'perturbative' particle. The actual collisional
width then is approximated by
\be
\label{cx} \Gamma_{coll}^i \approx \gamma_i \frac{\sum_j v_{ij}
\sigma_{ij}}{\sum_j} \, , \ee where the sum over $j$ runs over all
baryons in the local cell, $v_{ij}$ is the relative velocity in
the meson-baryon cms, $\gamma_i$ is the Lorentz-factor of the particle
with respect to the
local rest frame of the baryons and $\sigma_{ij}$ their total cross section
at invariant energy $\sqrt{s}_{ij}$. Note, that in (\ref{cx}) the
final state Pauli-blocking has been neglected for the baryon which
should be reasonable at the high bombarding energies of interest
here.

Apart from the description of particle propagation and
rescattering the results of the transport approach also depend on
the initial conditions, ${\vec X}_i(0), {\vec P}_i(0), M_i^2(0)$.
In view of nucleus-nucleus collisions, i.e. two nuclei impinging
towards each other with a laboratory momentum per particle
$P_{lab}/A$, the nuclei can be considered as in their respective
groundstate, which in the semiclassical limit is given by the
local Thomas-Fermi distribution \cite{CMMN}. Additionally the
virtual mass $M_i^2$ for nucleons has been determined by
Monte-Carlo according to the Breit-Wigner distribution
(\ref{alg_spectral}) assuming an in-medium width $\Gamma_0 $ = 1
MeV. We mention that varying $\Gamma_0$ from 1 -- 5 MeV does not
change the results to be presented in Section 3 within the
statistics achieved. For the vacuum width of stable hadrons we
have used $\Gamma_V $ = 1 MeV which implies that nucleons
propagating to the continuum in the final state of the reaction
achieve their vacuum mass on the 0.1 $\%$ level.

\section{Nucleus-nucleus collisions}
Our concrete applications we first carry out for nuclear reactions
at SIS energies (1 - 2 A GeV) that have been analysed within
conventional transport models to a large extent (cf. Ref.
\cite{CB99} and Refs. cited therein).

In view of Eq. (\ref{eomm}) we present for some randomly chosen
testparticles $i$ the off-mass-shell behaviour $M_i^2(t)-M_0^2$ as
a function of time in a central collision ($b$ = 1.5 fm) in Fig. 5
for $Au + Au$ at 1 A GeV. It is
seen that during the collision of the nuclei from $t \approx$ 7 --
25 fm/c the off-shellness of baryons reaches up to 0.8 GeV$^2$,
however, the nucleons become practically on-shell for $t \ge$ 35
fm/c. The individual sudden high mass excitations and subsequent
decays correspond essentially to $\Delta$ and N(1440) baryons.
Nucleons in their decay may be off-shell, but propagate again to
their on-shell mass in the continuum.

The baryon spectral function is shown in Fig. 6 as a function of
the invariant mass for the latter reaction at times of 0, 5, 10,
20, 40 and 60 fm/c. Apart from a broadening of the nucleon
spectral function at the initial time steps one observes that the
high mass tail is completely covered by the $\Delta$ and N(1440)
excitations. This is different from the results in \cite{Cass99}
at GANIL energies since in the latter case the $\Delta$ excitation
was dynamically suppressed and the high mass tail of the spectral
function dominated by nucleons. In fact, in 1 A GeV $Au + Au$
collisions the resonance high mass spectrum in the off-shell
calculations is only slightly enhanced as compared to the on-shell
calculations (without explicit representation).
Note that at t = 60 fm/c all resonances have decayed
and the nucleons have become on-shell again.

The dominance of the resonances for the high mass tail can also be
seen in the differential collision number $dN_{coll}/d\sqrt{s}$
for baryon-baryon collisions which is displayed in Fig. 7. Here
the dashed histogram stands for the on-shell result while the
solid histogram gives the off-shell distribution that extends well
below the two-nucleon threshold. The high energy tail in this
distribution dominantly arises from $N \Delta$ and $\Delta \Delta$
or $N N(1440)$ reactions, which are  similar in both calculations.
We thus find only a small enhancement in the high energy tail for
the off-shell case relative to the on-shell limit.

The latter observation can also be made in the transverse proton
momentum spectra $1/p_T d N_p/dp_T$ (Fig. 8) for $Au + Au$ at 1 A
GeV ($b$=1.5 fm) where the on-shell propagation (dashed histogram)
practically leads to the same result as the off-shell propagation
(solid histogram) except for the high momentum tail. Since the
elastic $NN$ cross section is taken here as a function of the
momentum difference in the cms (cf. Section 2.4) one might worry
if alternative prescriptions for $\sigma_{NN}$ could change the
results. In this context we have performed calculations using the
Cugnon parametrization $\sigma_{NN}(\sqrt{s})$ with $\sqrt{s}$
denoting the actual invariant energy of the off-shell nucleons and
adopting $\sigma_{NN} $ = 55 mb for invariant energies below 2
$m_N$, where $m_N$ denotes the nucleon vacuum mass. The transverse
momentum spectra according to this recipe (for the same reaction)
are shown in Fig. 8 in terms of the dot-dashed histogram, which
coincides with the 'default' off-shell result (solid histogram)
within the statistics.

\subsection{Meson production at SIS energies}

In principle also the pions should be
propagated with a dynamical spectral function since their coupling
to nucleons is very strong. In fact, pion collision rates in these
reactions lead to a collisional width which is much larger than
the pion mass itself. A straight forward selection of pion masses
(e.g. in the $\Delta$ decay) according to such a dynamical
spectral function leads to $M_\pi^2 < 0$ which implies that
such testparticles become acausal, i.e. move with velocities
$\beta >$ 1. In order to avoid such inconsistencies we treat
pions on-shell throughout this study. We note, however, that
microcausality is an essential issue that should also survive in
transport approximations. This is practically done by the explicit
requirement $M^2_i \ge 0$ in the transport calculation, but not
yet inherent in eqs. (\ref{eomr}-\ref{eome}).

Since the authors of Ref. \cite{Effe} claim a large enhancement in
the pion yield for $Au + Au$ at 1 A GeV and especially in the high
energy tails of the pion spectrum when describing  nucleon
off-shell propagation in their Monte-Carlo simulation, we show in
Fig. 9 (l.h.s.) the inclusive differential $\pi^+$  spectrum for
this system at $\theta_{lab} = 44 \pm 4^o$ within our off-shell
approach (solid histogram) in comparison to the on-shell limit
(dashed histogram) and the experimental data from the KaoS
collaboration \cite{Kaos97}. As in Ref. \cite{brat97} the $\pi^+$
spectrum is described quite well within the on-shell HSD approach.
As also seen from Fig. 9 there is only a very slight enhancement
of the high momentum tail in the pion spectrum for our off-shell
calculation which is still compatible with the experimental
spectrum from Ref. \cite{Kaos97} within the statistical errors.
Note that the fluctuations in the histograms with respect to an
average exponential spectrum provide some information about the
statistical accuracy. Thus our off-shell approach -- based on the
Kadanoff-Baym equation (\ref{trans_approx}) -- is not in conflict
with the experimental data contrary to the model from Ref.
\cite{Effe}.

Since kaons couple only weakly to nucleons and are not absorbed at
low energies their collisional width is rather small such that
they may be treated on-shell to a good approximation. In their
differential production cross section we thus only can test the
effects from the off-shell propagation of baryons. The inclusive
$K^+$ spectra at $\theta_{lab} = 44 \pm 4^o$ for $Au + Au$ at 1 A
GeV are shown in Fig. 9 (r.h.s.) where we compare our off-shell
results (solid histogram) with the on-shell limit (dashed
histogram) and the experimental data from Ref. \cite{Misko} (full
circles) and Refs. \cite{SengerP,kaos1} (open circles) where the
latter differ on average by a factor of 2. The off-shell
calculations give a slightly higher $K^+$ yield than the on-shell
calculations, however, are still well within the error bars of the
presently available data. We mention that we have performed the
calculations without any kaon potential; a slightly repulsive kaon
potential will drop the $K^+$ yield and harden the spectrum
slightly as discussed in Ref. \cite{brat97}.

The relative increase of the $K^+$ spectrum by about 50\% in the
off-shell calculation with respect to the on-shell result should
also be compared with the relative sensitivity to the
incompressibility $K$ of the nuclear equation-of-state (EoS)
\cite{Aich75}. According to the early relativistic calculations by
Lang et al. \cite{Lang} the inclusive $K^+$ yield in $Au + Au$
collisions at 1 A GeV is enhanced by about a factor of 2 when
decreasing the incompressibility $K$ from 380 MeV to 200 MeV. When
restricting to incompressibilities 200 MeV $\leq K \leq$ 300 MeV
this relative sensitivity reduces to a similar change in the $K^+$
yield for $Au + Au$ at 1 A GeV as obtained from the off-shell
dynamics relative to the on-shell treatment. It thus appears
questionable if the incompressibility of the EoS can be determined
from the systematics of $K^+$ spectra alone since also the $K^+$
in-medium potential potential is not yet determined very
accurately.

We continue with pion production in $Ni + Ni$ collisions at 1.8 A
GeV since for this system also $K^+$ and $K^-$ spectra have been
measured \cite{SengerP,Kaos97,kaos1}. The inclusive differential
cross section for $\pi^+$ from $Ni + Ni$ collisions at 1.8 A GeV
is shown in Fig. 10 as a function of the pion momentum in the cms
in comparison to the data of the KaoS Collaboration taken at
$\Theta_{lab} = 44 \pm 4^0$ \cite{SengerP,kaos1}. The calculation
with an off-shell propagation of baryons is represented by the
solid histogram while the on-shell limit is displayed in terms of
the dashed histogram. Within the numerical accuracy both results
almost coincide which in view of Fig. 7 is not an exciting
surprise. Both results, furthermore, are in a reasonable agreement
with the measured spectrum (full squares) which suggests that no
'unphysical' approximations have been introduced within the
'recipies' for the off-shell transition cross sections.

As might have been anticipated from the studies before, also the
inclusive $K^+$ spectra from $Ni + Ni$ at 1.8 A GeV -- plotted as
a function of the kaon momentum in the cms -- is only weakly
sensitive to the off-shell propagation of baryons as shown in Fig.
11 since the differential $\sqrt{s}$ distribution for
baryon-baryon collisions does not differ very much and also the
pion-baryon production channel is similar (except for the high
momentum tail). The latter fact one might also extract from the
low sensitivity of the pion spectrum to an off-shell propagation
in Fig. 10 for this reaction. We note that the kaons again have
been propagated without any in-medium potential, which should be
slightly repulsive in line with Refs.
\cite{Kaplan,Weise,Schaffner}. As shown in Ref. \cite{brat97} such
a repulsive potential will suppress the kaon spectra especially at
low momenta. Since we have employed the same production cross
sections as in Ref. \cite{brat97} we again reproduce the data from
the KaoS Collaboration, taken at $\Theta_{lab} = 44 \pm 4^0$, best
without a kaon potential, where the off-shell calculation seems to
be in even better agreement with the data. We mention that the
'theoretical' error bars in Fig. 11 are of statistical nature only
and computed as $\pm 0.5 S_p/\sqrt{N_p}$ where $S_p$ is the
calculated spectral point in the actual momentum bin and $N_p$
denotes the number of events contributing to this momentum bin.
Though the error bars between the on-shell (open triangles) and
off-shell calculation (full triangles) almost overlap, we point
out that the off-shell calculation gives a slightly harder
spectrum.

Note that the production channel $N\Delta \rightarrow N K^+ Y$,
where $Y$ denotes a hyperon, as well as the $\Delta \Delta
\rightarrow K^+ N Y$ channel is not known experimentally and
simple isospin factors as extracted from pion exchange
\cite{brat97} might not be appropriate. Though there are some
recent efforts to resolve this uncertainty within extended boson
exchange models \cite{Sib}, the latter models will hardly be
tested experimentally. This general uncertainty has to be kept in
mind when comparing transport calculations to experimental kaon
spectra.

We step on with the production of antikaons in the same reaction.
Antikaons couple strongly to nucleons and thus achieve a large
collisional width in the nuclear medium. The calculations of Ref.
\cite{Sibkaon}, which are based on a dispersion approach, give a
collisional width of about 100 MeV of $K^-$ mesons at moderate
momenta and nuclear matter density $\rho_0$. Thus off-shell
antikaons might be produced at far subthreshold energies (cf. Fig.
4), become asymptotically on-shell and thus enhance the $K^-$
yield. Note, that this mechanism also might explain the $K^-$
enhancement seen by the FRS, KaoS and FOPI collaborations
\cite{kaos1,KaoS,FRS} in $Ni + Ni$ reactions around 1.8 A GeV.

Within our present 'modelling' of off-shell production processes,
which are governed by phase-space close to threshold energies, we
find an enhancement by about a factor $\sim$ 2 in the production
of antikaons when treating baryons as well as antikaons off-shell.
This is demonstrated in Fig. 12 where we show the $K^-$ spectra as
a function of their momentum in the cms for the off-shell (solid
histogram, full triangles) and on-shell propagation (dashed
histogram, open triangles). As in case of Fig. 11 the
'theoretical' error bars are statistical, only, and taken as $\pm
0.5 S_p/\sqrt{N_p}$ where $S_p$ and $N_p$ denote the actual value
in the momentum bin and number of events, respectively. Both
results underestimate the data from the FRS and KaoS
Collaborations \cite{SengerP,kaos1,FRS} such that the conclusion
in Refs. \cite{Cass97,lix} on attractive $K^-$ self energies in
the nuclear medium persists, though the off-shell calculations
suggest somewhat smaller $K^-$ potentials. We mention that the
dispersion analysis of the $K^-$ potential in nuclear matter in
Ref. \cite{Sibkaon} also yields somewhat smaller antikaon
potentials than that extracted in Refs. \cite{Cass97,lix} which is
in line with our present finding. However, in order to obtain more
model independent results on the antikaon self energy, precise
data on antikaon flow from nucleus-nucleus collisions are urgently
needed.

\subsection{AGS energies}
At AGS energies of $\approx $ 11 A GeV the invariant energy
$\sqrt{s}$ of the initial nucleon-nucleon collisions is about 5
GeV which implies that the nucleons are excited to continuum
states -- denoted by {\it strings} -- which decay according to
phase-space \cite{LUND} and fixed quark/diquark or $s/(u,d)$
ratios after a {\it formation time} $\tau_F$ = 0.8 fm/c. Some
details of the decay scheme are given in Ref. \cite{Geiss}. The
FRITIOF 7.2 version - as implemented in the HSD transport approach
-  also includes the production of unstable particles of width
$\Gamma_h$ by Monte-Carlo. The width $\Gamma_h$ has been modified
dynamically according to the actual total width at space-time
point $X$ which allows to simulate the production of hadrons with
broad spectral functions from string decay in a straight forward
manner. The low energy baryon-baryon and meson-baryon reactions
are treated as for SIS energies (cf. Section 2.4).

We here present only a single study for central $Au + Au$
reactions at 11.3 A GeV ($b \leq$ 3.5 fm) and concentrate on the
rapidity and transverse mass spectra of pions, kaons and
antikaons. The results of our calculations are displayed in Figs.
13 and 14 for the off-shell calculation (solid histograms) and the
on-shell calculations (dashed histograms) in comparison to the
data from Refs. \cite{EOS1,E866}. Since at this bombarding energy
the dynamics are dominated by continuum string excitations, a
broadening of the spectral functions is found to play a minor role
here. Note, that the $K^{\pm}$ spectra for $Au + Au$ collisions at
AGS energies are underestimated in comparison to the data, which
in Ref. \cite{Geiss} has been attributed to nonhadronic degrees of
freedom  or a partial restoration of chiral symmetry during the
high density collision phase.

\section{Summary}
In this work we have employed the semiclassical off-shell
transport approach from Ref. \cite{Cass99}, that in first order in
the gradient expansion describes the virtual propagation of
particles in the invariant mass squared $M^2$ besides the
conventional propagation in the mean-field potential (given by the
real part of the retarded self energy), to analyse nucleus-nucleus
collisions at SIS and AGS energies. Note that in conventional
transport approaches the imaginary part of the self energy is
reformulated in terms of a collision integral and simulated by
on-shell binary collisions, only. We here additionally account for
the off-shell propagation of particles due to the imaginary part
of the self energy in eqs. (\ref{eomr},\ref{eomp},\ref{eomm})
which describe the dynamical evolution of the particle spectral
function. The additional imaginary part of the self energy (or
local collision rate $\Gamma_{coll}$) is determined
by the collision integrals themselves and can be used in transport
approaches without introducing any new assumptions or parameters
provided that the off-shell transition amplitudes $T$ are known in
the collision integral (\ref{Icoll}).
In addition to Ref. \cite{Cass99} we have
included momentum-dependent self energies for baryons and mesons
in our present approach. However, a final answer to the role of
off-shell hadrons in nucleus-nucleus collisions we still have to
leave for future due to the 'ad hoc' assumptions involved in the
treatment of the off-shell transition probabilities.

We have presented dynamical calculations of the novel transport
theory for nucleus-nucleus collisions at SIS and AGS energies
where we can test its results in comparison to experimental data.
We find that the off-shell propagation of nucleons practically
does not change the rapidity distributions $dN/dy$ and has only a
minor effect on the transverse momentum spectra of protons within
the statistics reached except for the very high momentum tails.
The distribution of baryon-baryon collisions in the invariant
energy $\sqrt{s}$ is found to be also enhanced only for high
invariant energies since here the collisions with or between
resonances -- which are only slightly affected in their high mass
spectrum -- dominate the spectrum. Again except for high momentum
tails we find no dramatic change in the pion and $K^+$ spectra at
SIS energies for $Au + Au$ at 1.0 A GeV and $Ni + Ni$ at 1.8 A
GeV, our results being well in line with the data of the KaoS
Collaboration. This no longer holds for the $K^-$ spectra from $Ni
+ Ni$ collisions at 1.8 A GeV which are enhanced by a factor of
$\sim$ 2 relative to the on-shell calculation within the
statistics reached. We attribute this enhancement to a broad
spectral function of antikaons at high baryon density and to the
'subthreshold' energy of 1.8 A GeV considered. However, even in
case of the off-shell propagation of baryons and mesons the
experimental $K^-$ spectra are underestimated and attractive
antikaon potentials are still needed to achieve a proper
description.

At AGS energies ($\approx $ 11 A GeV) the particle production
in the HSD approach essentially occurs via the
excitation and decay of strings which
can be viewed as continuum excitations of hadrons. Any spectral
broadening of the 'continuum' thus is not likely to be seen in the
asymptotic particle spectra of pions, kaons or antikaons
especially since they are most abundantly produced far above the
individual $NN$ or $\pi N$ thresholds.

We finally point out that although our off-shell transport
approach appears to be in a reasonable agreement with the
differential experimental data at least for SIS energies (except
for $K^-$ when discarding antikaon potentials), the question of
proper off-shell transition amplitudes in the collision terms
remains an open problem that has to addressed in the near future.
Some steps in this direction e.g. have been taken in Ref.
\cite{Bozek}.

\vspace{1cm} The authors like thank E. L. Bratkovskaya, C. Greiner and S.
Leupold\footnote{After completion of this work an independent
preprint appeared by the author \cite{Leupold} in which the
testparticle equations of motion (18)-(20) have been rederived in
the nonrelativistic limit.} for stimulating discussions
throughout this study. Furthermore, they
also acknowledge lively, though controversal, exchange of arguments with
M. Effenberger and U. Mosel.

%----------------------------------------------------------------------

%\end{document}

%----------------------------------------------------------------------
%
%
%
\newpage
%
%
%
% figure 1
\begin{figure}[h]
\phantom{a}\vspace*{-1cm}
\epsfig{file=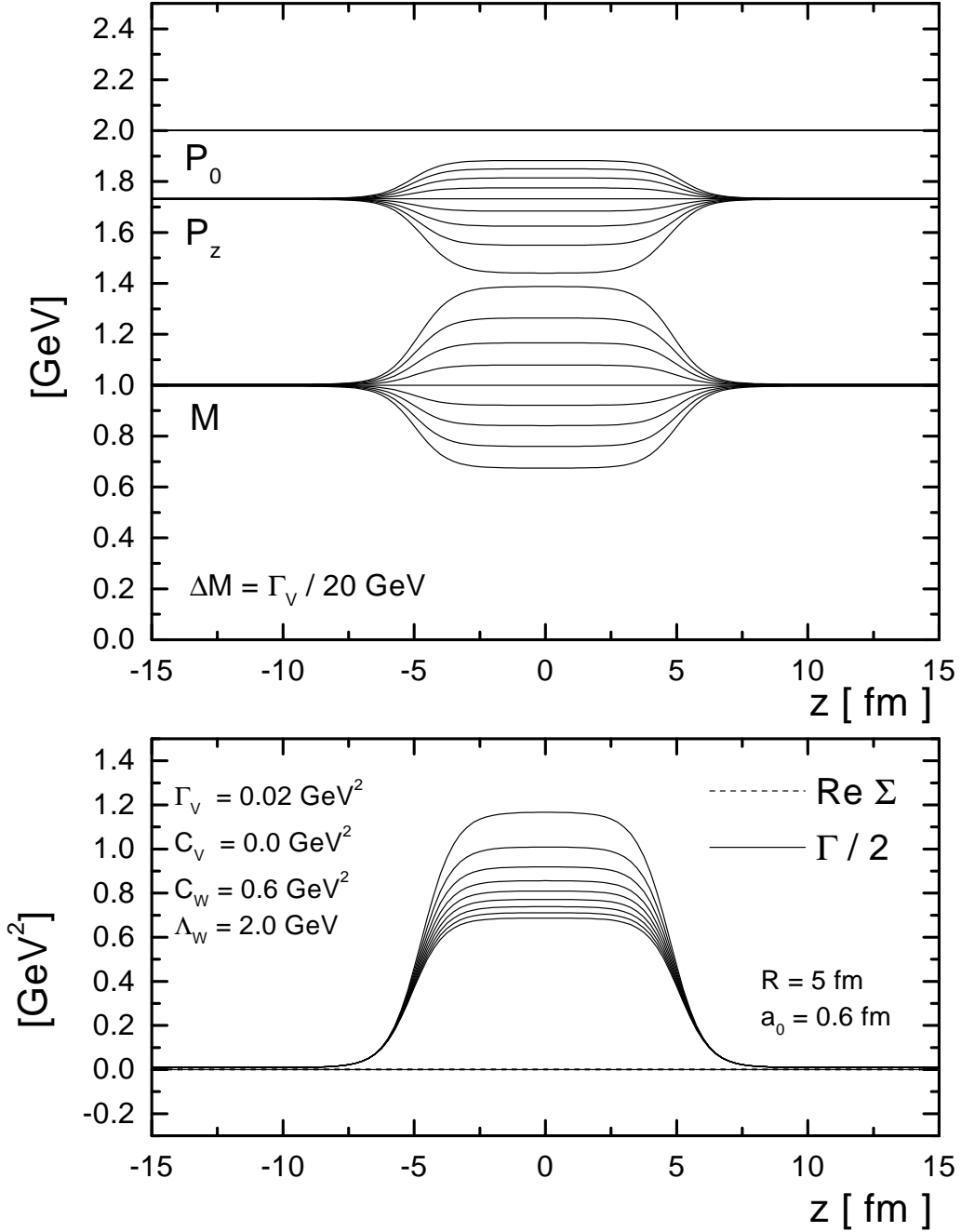,width=15cm}
\vspace*{-2cm} \caption{upper part: $P_{i0}$, $P_{iz}$ and $M_i$
as a function of $z(t)$ for a momentum-dependent imaginary
potential with $C_W = 0.6$ GeV$^2$ and $\Lambda_W = 2.0$ GeV
(lower part). The vacuum width is chosen as $\Gamma_V = 0.02$
GeV$^2$ and the initial separation in the mass parameter of the
testparticles is $\Delta M = \Gamma_V / 20$.} \label{fig1m}
\end{figure}
%
%
% figure 2
\begin{figure}[h]
\phantom{a}\vspace*{-1cm}
\epsfig{file=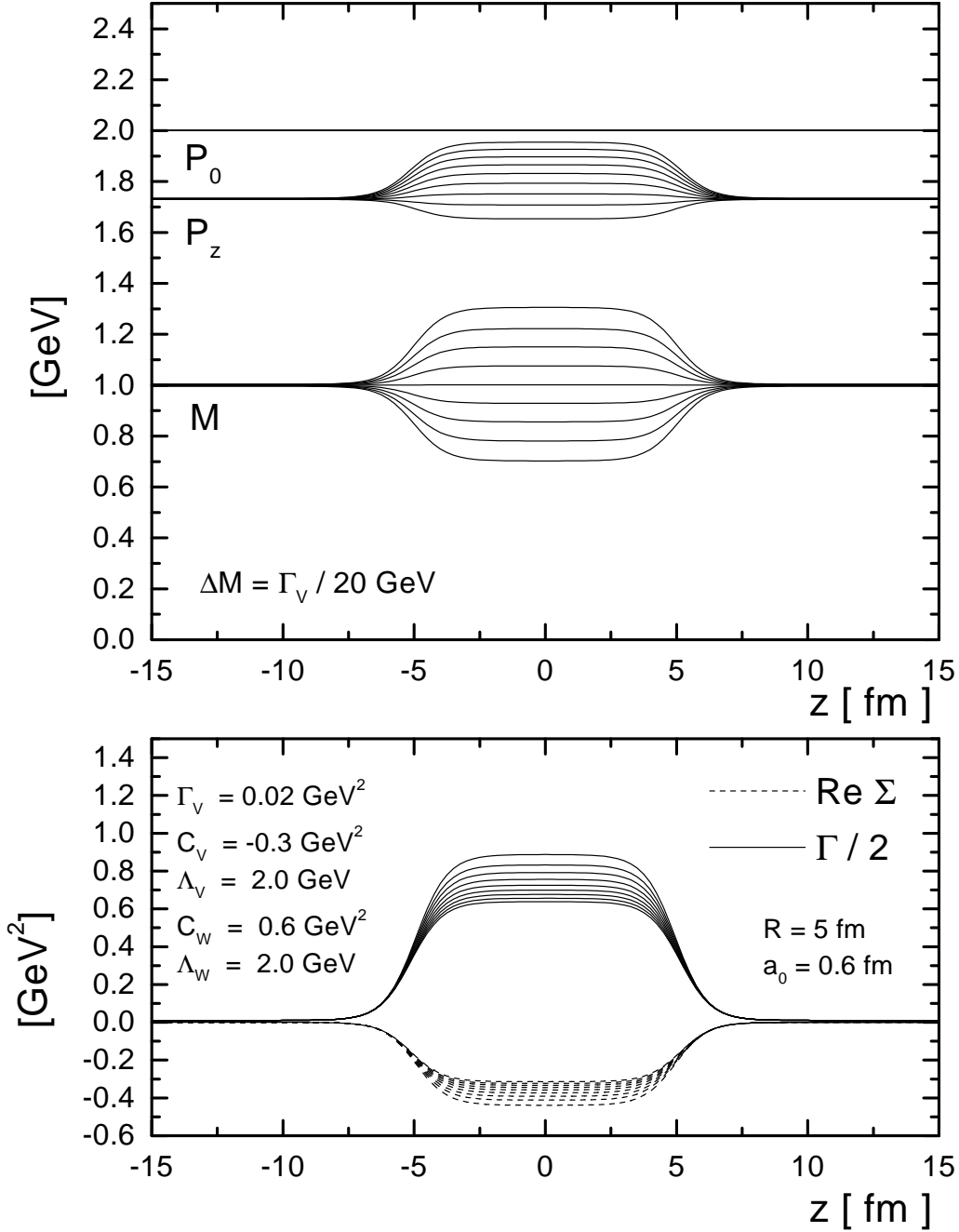,width=15cm}
\vspace*{-2cm}
\caption{upper part: $P_{i0}$, $P_{iz}$ and $M_i$ as a function of
$z(t)$ for a momentum-dependent complex potential with
$C_V = -0.3$ GeV$^2$, $\Lambda_V = 2.0$ GeV,
$C_W = 0.6$ GeV$^2$ and $\Lambda_W = 2.0$ GeV (lower part).
For the vacuum width and the initial mass separation the same
values are used as in Fig. 1.}
\label{fig2m}
\end{figure}
%
%
% figure 3
\begin{figure}[h]
\phantom{a}\vspace*{-2cm}
\epsfig{file=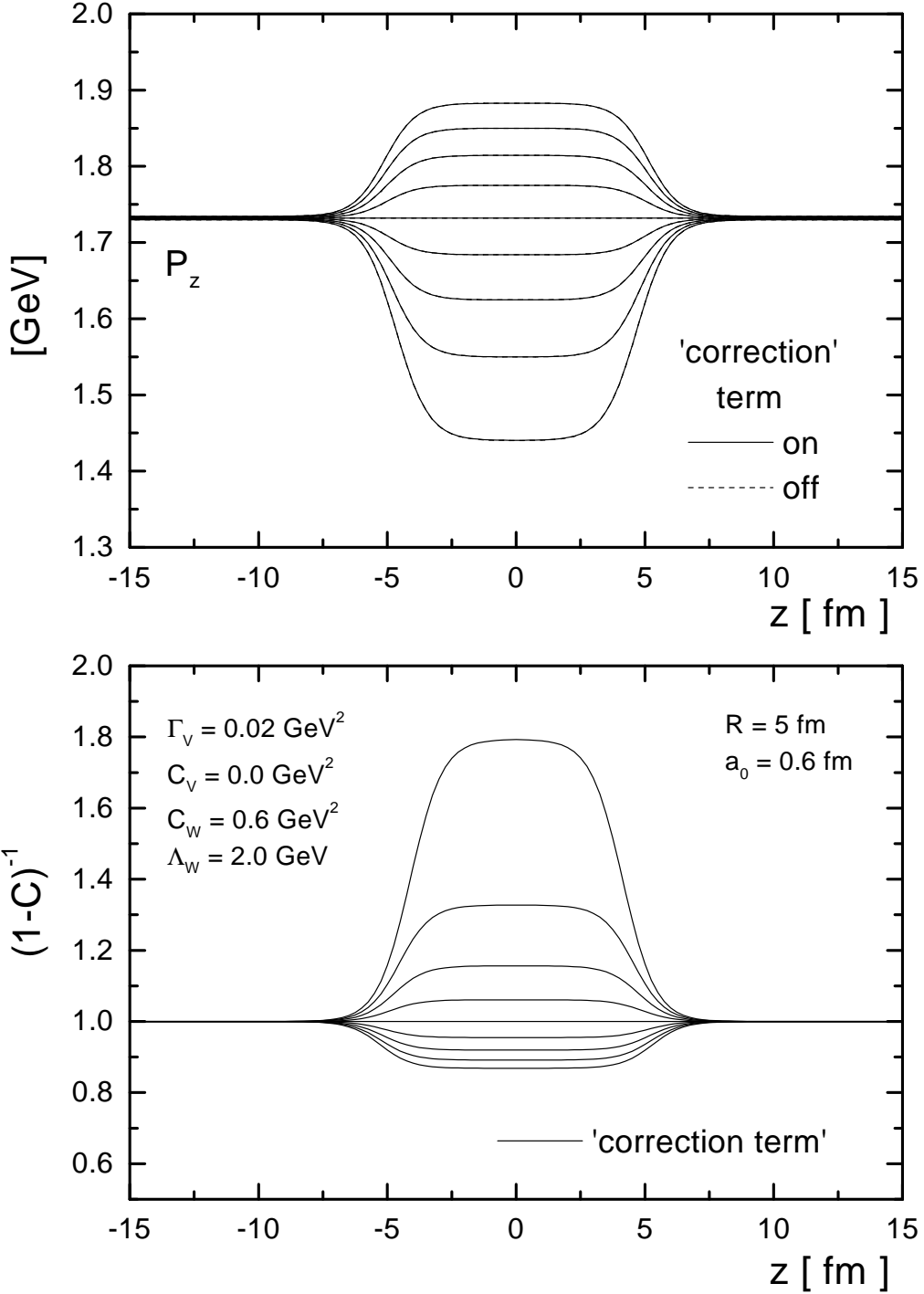,width=15cm}
\vspace*{-1cm}
\caption{lower part: Correction term
$(1-C_{(i)})^{-1}$ as a function of $z(t)$ for the same imaginary
potential as in Fig. 1. Upper part: $P_{iz}$ as a function of
$z(t)$ with and without including of the correction term. Both
curves cannot be resolved separately within the line width.}
\label{fig3m}
\end{figure}
%
%
% figure 4
\begin{figure}[h]
\phantom{a}\vspace*{-1cm}
\epsfig{file=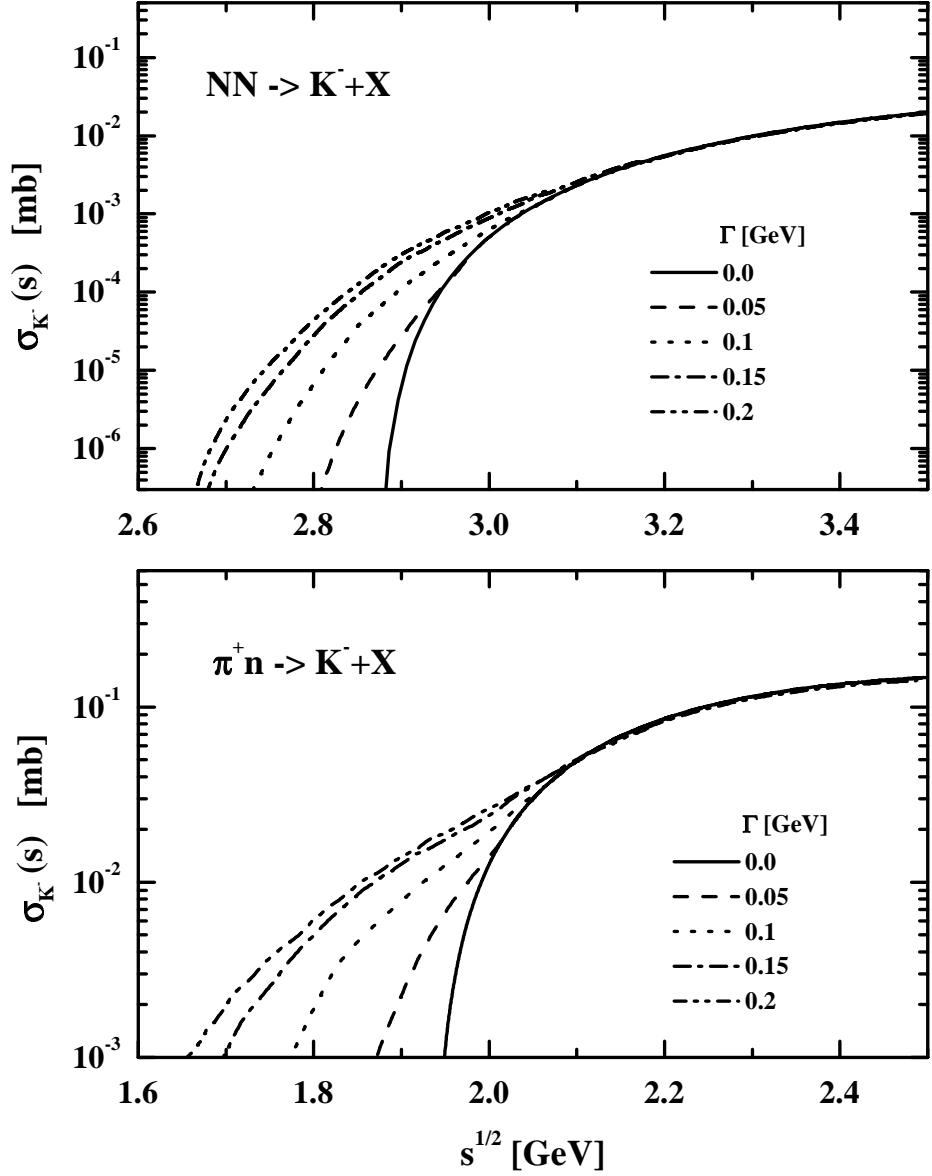,width=15cm}
\vspace*{-3cm}
\caption{The $K^-$ cross section from $NN$ (upper part) and $\pi N$
(lower part) collisions as a function of the invariant energy
$\sqrt{s}$ for different collisional width $\Gamma$ of the antikaon
spectral function in the medium according to the model discussed in
the text.}
\label{fig1}
\end{figure}
%
%
% figure 5
\begin{figure}[h]
\phantom{a}\vspace*{-1cm}
\epsfig{file=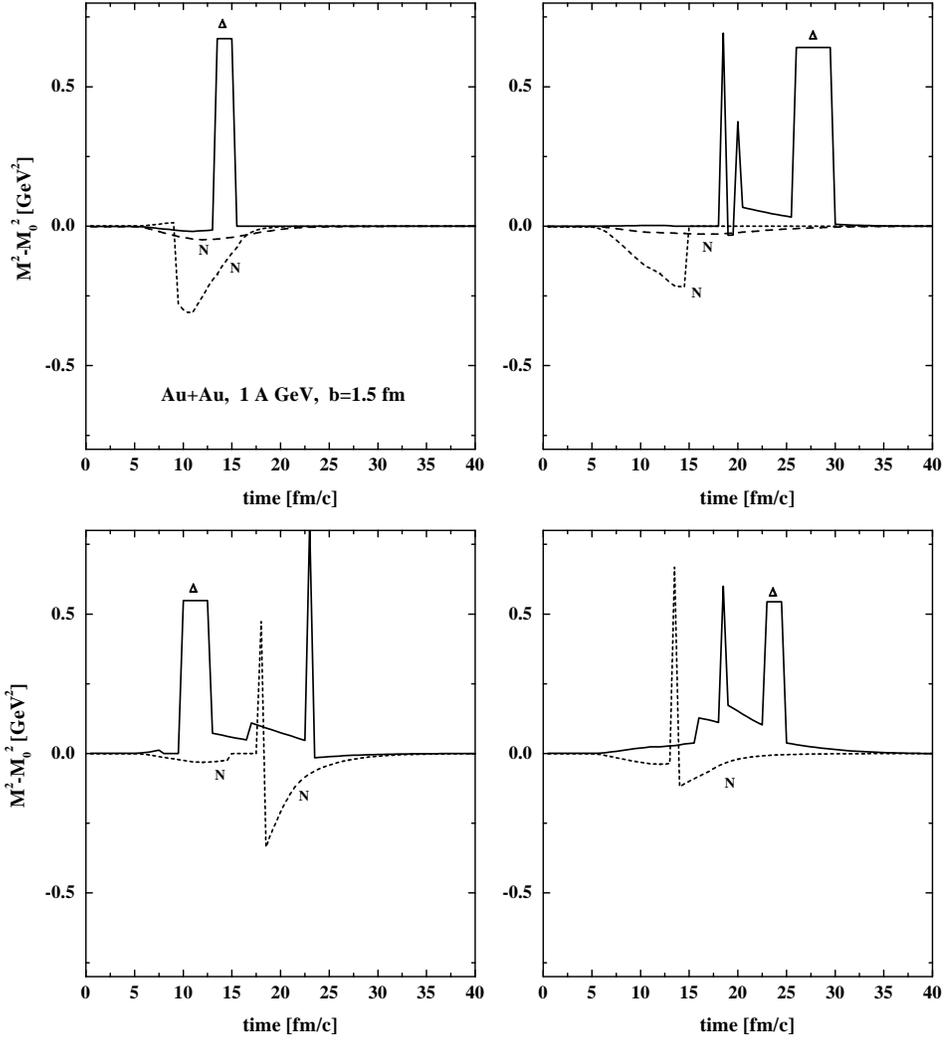,width=15cm}
\vspace*{-3cm}
\caption{Some randomly chosen examples for the baryon off-shell propagation
in mass in $Au + Au$ collisions at 1 A GeV and $b$ = 1.5 fm. The sudden spikes
correspond to $\Delta$ or $N^*$ excitations.}
\label{fig2}
\end{figure}
%
%
% figure 6
\begin{figure}[h]
\phantom{a}\vspace*{-2cm}
\epsfig{file=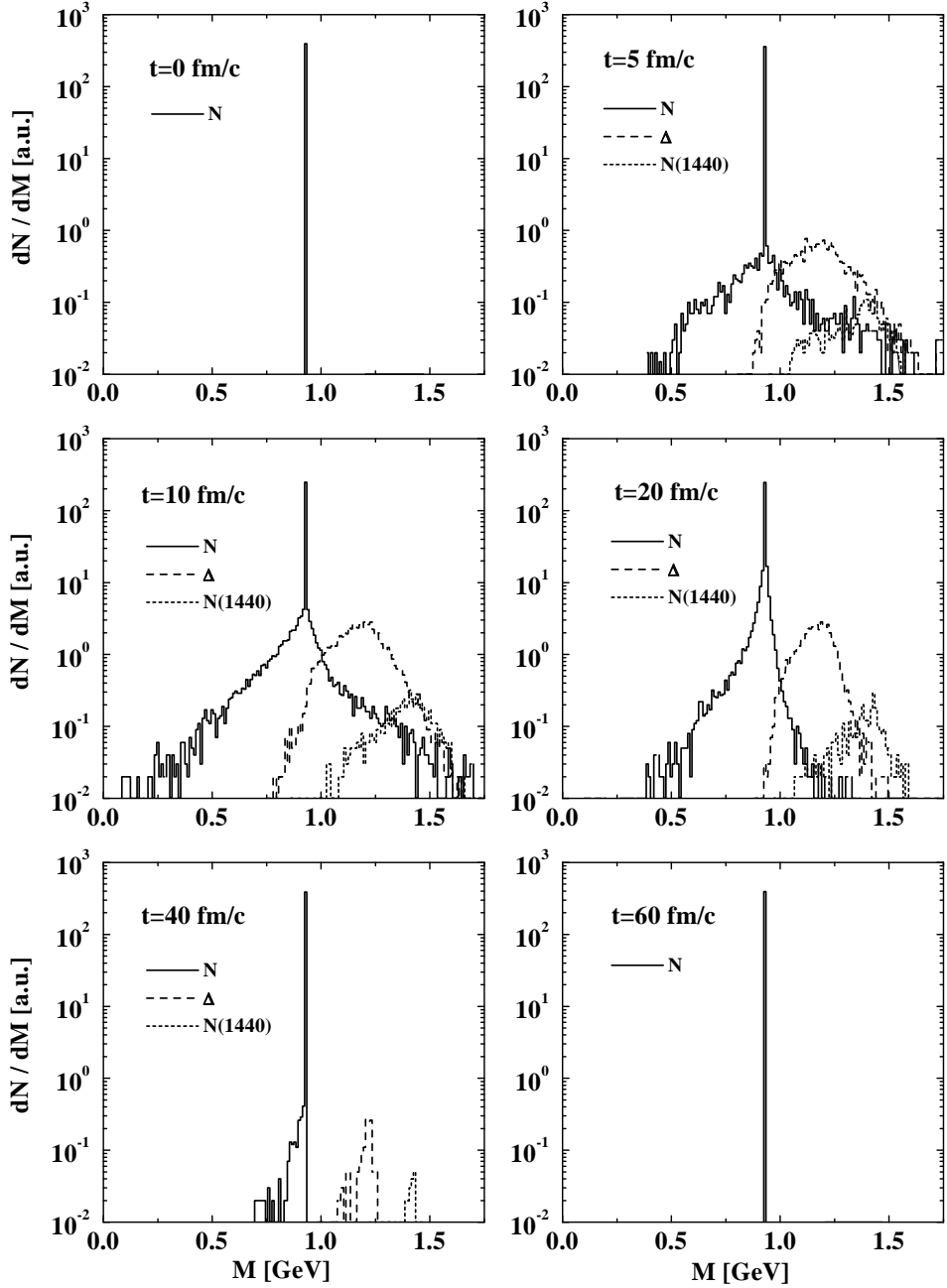,width=15 cm}
\vspace*{-2.0cm}
\caption{The baryon distribution in mass $M$ for $Au + Au$ at 1 A GeV and
$b$ = 1.5 fm for times of 0, 5, 10, 20, 40 and 60 fm/c. The dashed and dotted
lines stand for $\Delta$ and $N(1440)$ resonances, respectively.}
\label{fig3}
\end{figure}
%
%
% figure 7
\begin{figure}[h]
\phantom{a}\vspace*{-1cm}
\epsfig{file=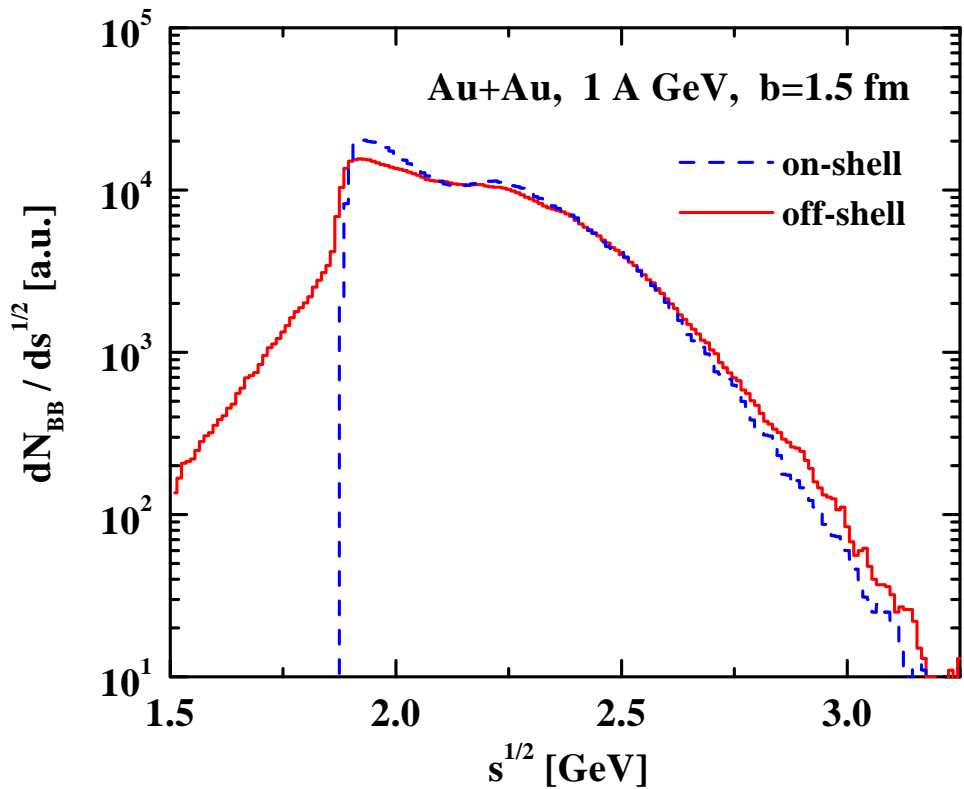,width=15cm}
\vspace*{-4cm}
\caption{The number of baryon-baryon (BB) collisions as a function of
the invariant energy $\sqrt{s}$ for $Au + Au$ at 1 A GeV and $b$ = 1.5
fm. The solid line is obtained from including the off-shell propagation
of baryons in the transport approach while the dashed line stands for
the result in the on-shell limit.}
\label{fig4}
\end{figure}
%
%
% figure 8
\begin{figure}[h]
\phantom{a}\vspace*{-1cm} \epsfig{file=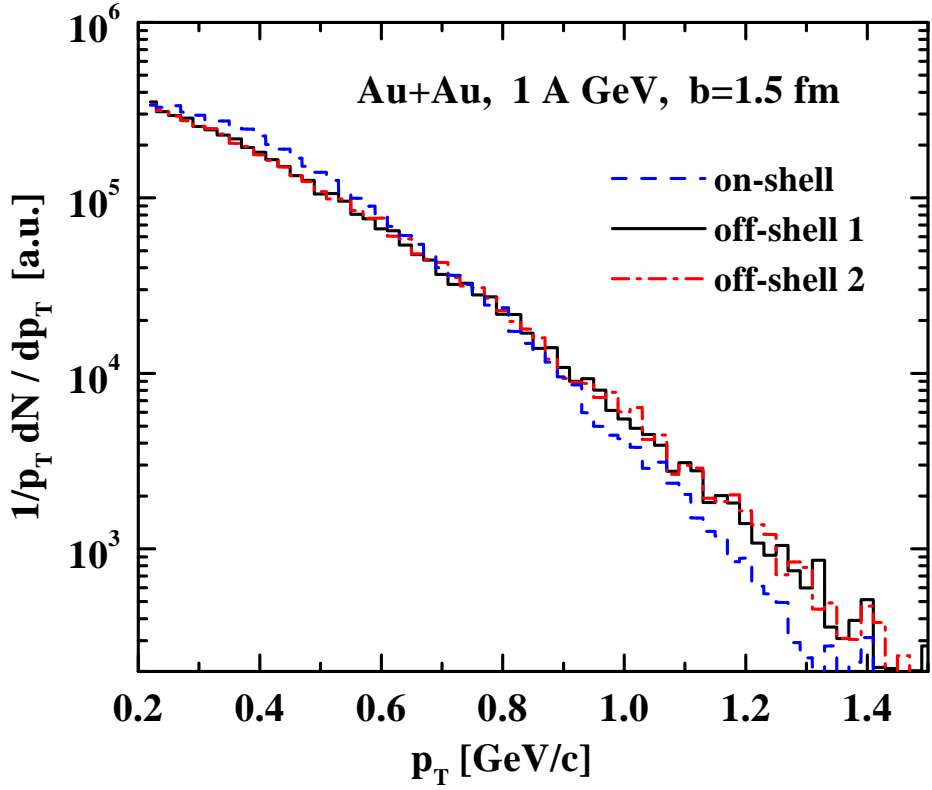,width=15cm}
\vspace*{-4cm} \caption{The transverse momentum spectra of protons
$1/p_T d N_p/d p_T$ for $Au + Au$ at 1 A GeV and impact parameter
$b$ = 1.5 fm. The dashed histogram is the result from the on-shell
propagation while the solid histogram is obtained including the
'default' off-shell propagation of baryons. The dot-dashed
histogram displays the result for an alternative modelling of the
'off-shell' elastic $NN$ collisions (see text); the two off-shell
calculations give the same $p_T$ spectra within the statistical
accuracy. } \label{fig5}
\end{figure}
%
%
% figure 9
\begin{figure}[h]
\phantom{a}\vspace*{-1cm} \epsfig{file=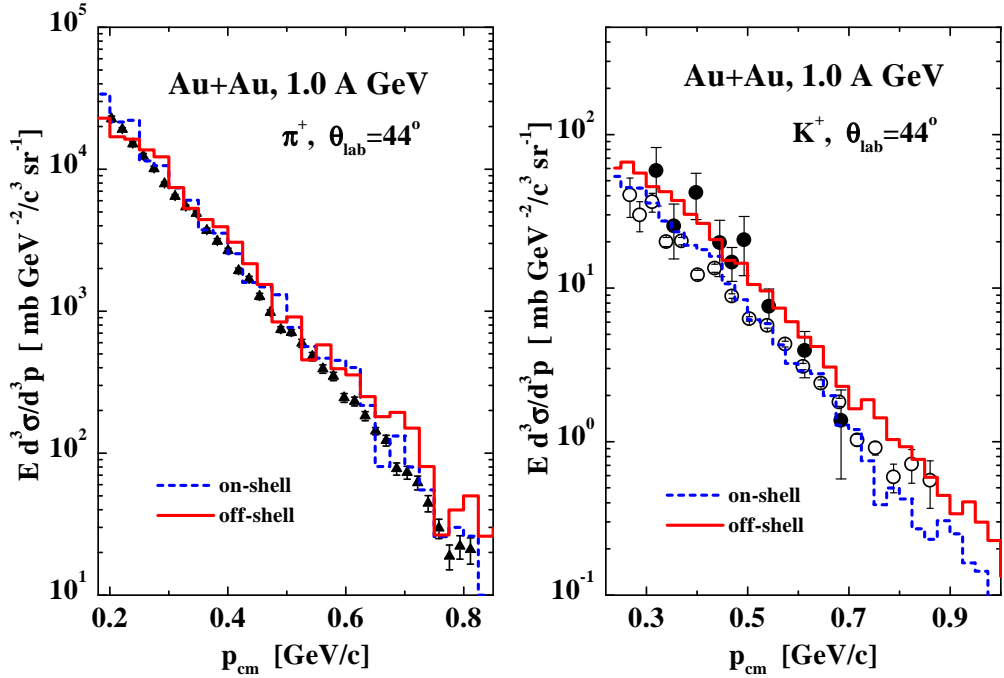,width=15cm}
\vspace*{-6cm} \caption{(l.h.s.): The inclusive  momentum spectra
of positive pions at $\theta_{lab} = 44 \pm 4^o$ for $Au + Au$ at
1 A GeV in comparison to the experimental data from the KaoS
Collaboration \protect\cite{Kaos97} (full triangles) displayed as
a function of the momentum in the nucleus-nucleus cms. (r.h.s.):
The inclusive momentum spectra of positive kaons at $\theta_{lab}
= 44 \pm 4^o$ for $Au + Au$ at 1 A GeV in comparison to the
experimental data from the KaoS Collaboration (full circles
\protect\cite{Misko}, open circles \protect\cite{SengerP}). The
dashed histograms show the results from the on-shell propagation
while the solid histograms are obtained including the off-shell
propagation of baryons. The fluctuations of the histograms with
respect to an average exponential spectrum provide an estimate for
the statistical error bars of the calculations. } \label{fig5new}
\end{figure}
%
%
% figure 10
\begin{figure}[h]
\phantom{a}\vspace*{-1cm} \epsfig{file=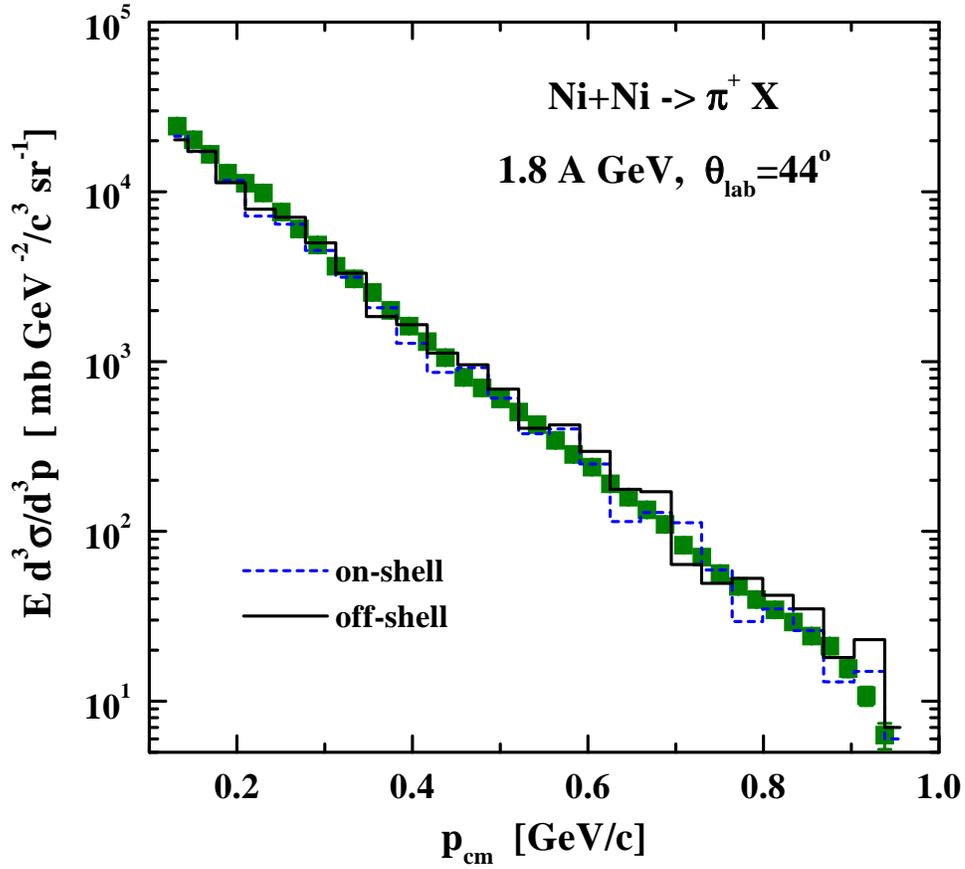,width=15cm}
\vspace*{-4cm} \caption{The inclusive spectra of positive pions
for $Ni + Ni$ at 1.8 A GeV including the off-shell propagation in
the transport approach (solid histogram) and in the on-shell limit
(dashed histogram) in comparison to the experimental data from
\protect\cite{kaos1} (full squares) at $\theta_{lab}$ = 44$^o$.}
\label{fig6}
\end{figure}
%
%
% figure 11
\begin{figure}[h]
\phantom{a}\vspace*{-1cm} \epsfig{file=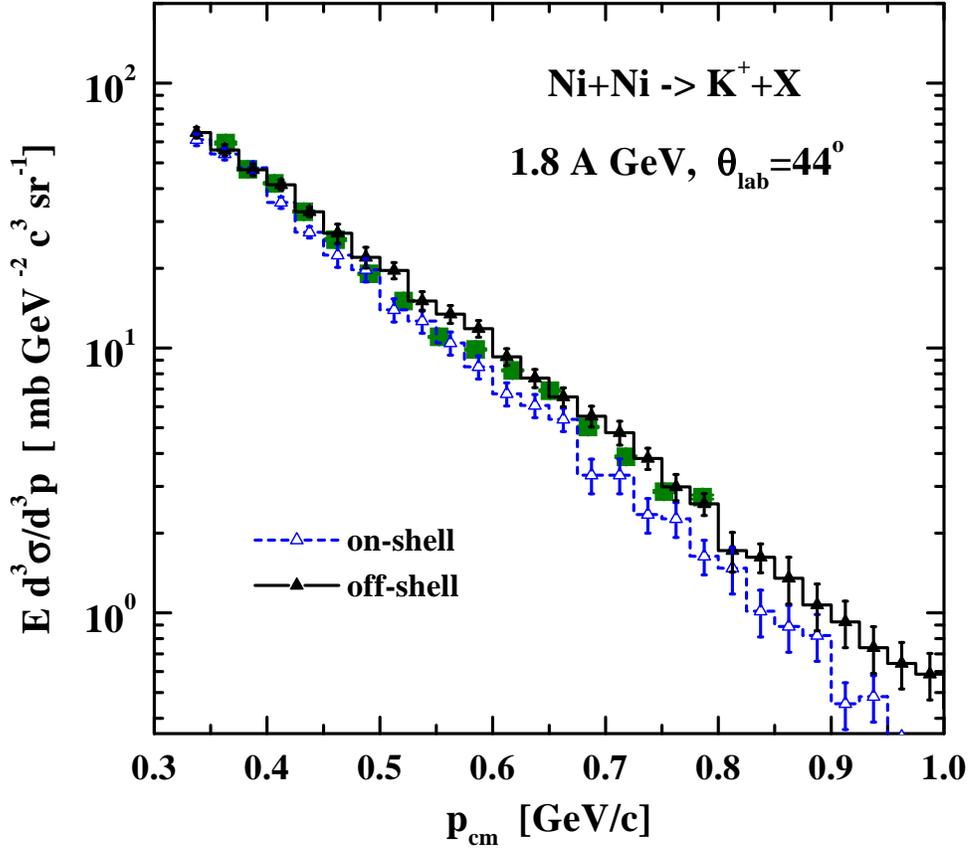,width=15cm}
\vspace*{-4cm} \caption{The inclusive spectra of positive kaons
for $Ni + Ni$ at 1.8 A GeV including the off-shell propagation in
the transport approach (solid histogram, full triangles) and in
the on-shell limit (dashed histogram, open triangles) in
comparison to the experimental data from \protect\cite{kaos1}
(full squares) at $\theta_{lab}$ = 44$^o$. The 'theoretical' error
bars are statistical, only (see text). } \label{fig7}
\end{figure}
%
%
% figure 12
\begin{figure}[h]
\phantom{a}\vspace*{-1cm}
\epsfig{file=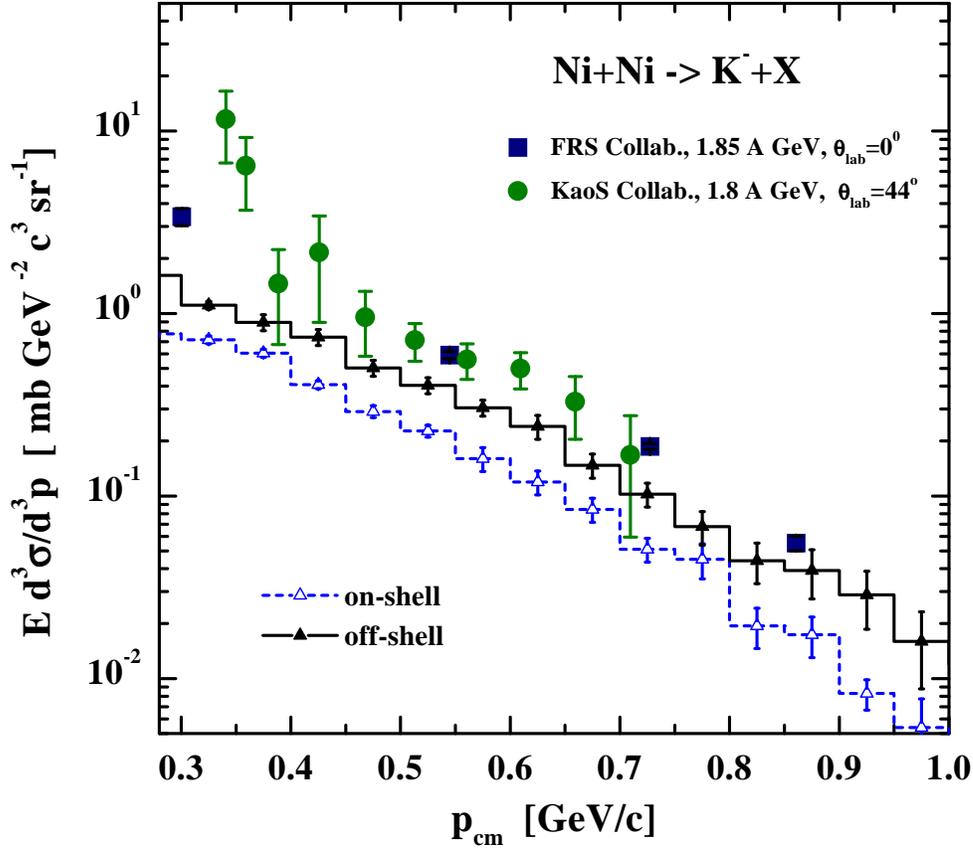,width=15cm}
\vspace*{-4cm}
\caption{The inclusive spectra of $K^-$ for $Ni + Ni$
 at 1.8 A GeV including the off-shell propagation in the
transport approach (solid histogram, full triangles) and in the
on-shell limit (dashed histogram, open triangles) in comparison to
the experimental data from Refs. \protect\cite{kaos1} (open
circles) and \protect\cite{FRS} (full squares) at $\theta_{lab}$ =
44$^o$ and $\theta_{lab} $ = 0$^0$, respectively. Note that no
antikaon potentials have been included in the calculations. The
'theoretical' error bars are statistical, only (see text).}
\label{fig8}
\end{figure}
%
%
% figure 13
\begin{figure}[h]
\phantom{a}\vspace*{-1cm}
\epsfig{file=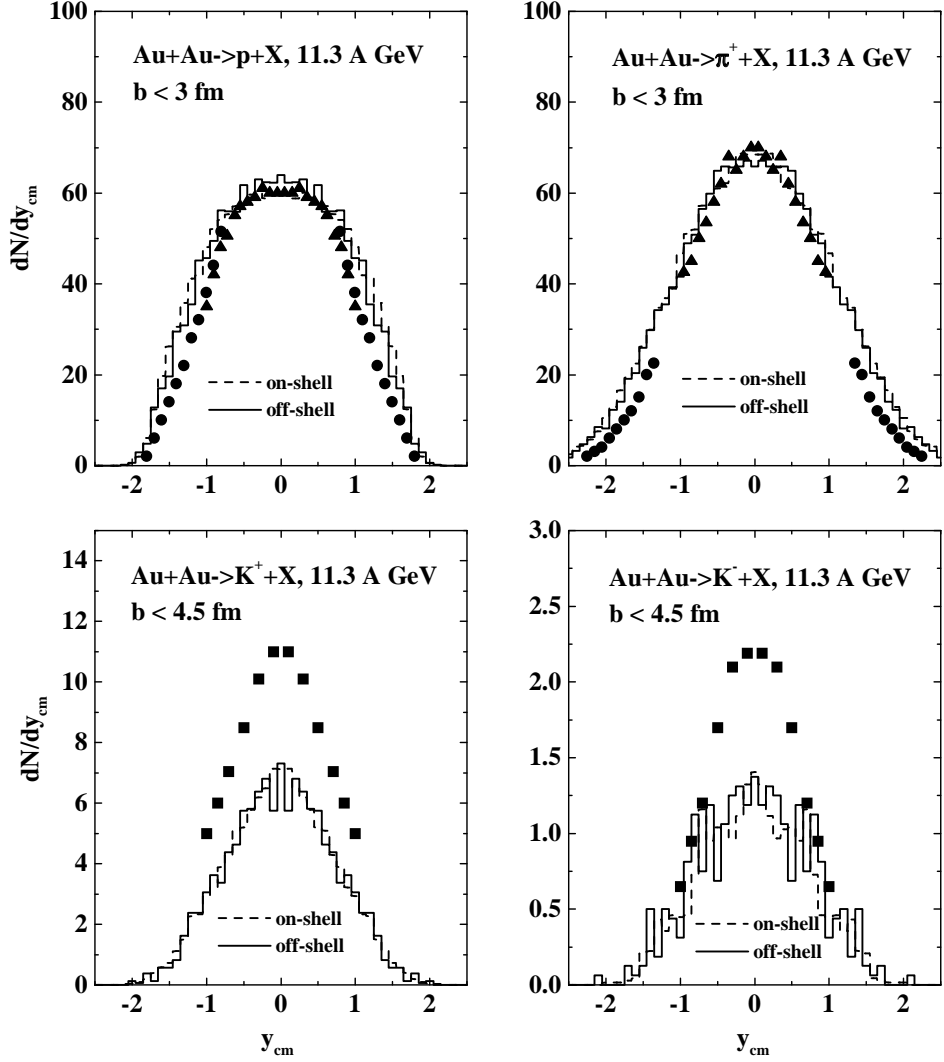,width=15cm}
\vspace*{-4cm} \caption{The rapidity distributions of protons,
$\pi^+, K^+$ and $K^-$ for central collisions of $Au + Au$ at 11.3
A GeV in comparison to the data from Refs.
\protect\cite{EOS1,E866}. The solid histograms are obtained
including the off-shell propagation in the transport approach
while the dashed histograms result from the on-shell limit.}
\label{fig9}
\end{figure}
%
%
% figure 14
\begin{figure}[h]
\phantom{a}\vspace*{-1cm}
\epsfig{file=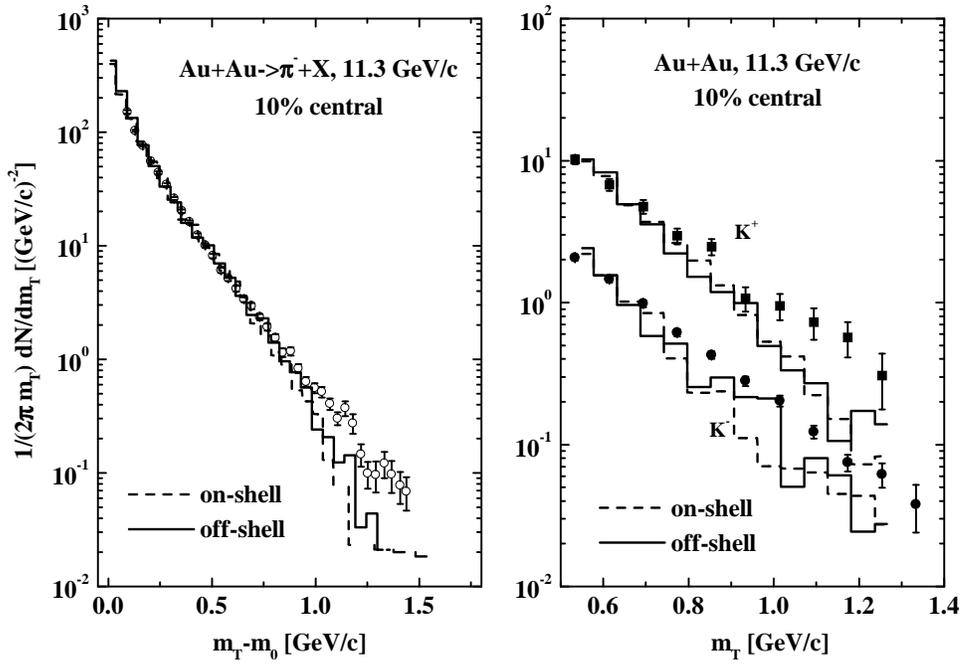,width=15cm}
\vspace*{-4cm}
\caption{The transverse mass spectra of $\pi^+$ (l.h.s.),
$K^+$ and $K^-$ mesons (r.h.s.) for central collisions of $Au + Au$
at 11.3 A GeV in comparison to the data from Refs. \cite{E866}.
The solid histograms are obtained including the off-shell propagation in the
transport approach while the dashed histograms result from the on-shell
limit.}
\label{fig10}
\end{figure}
\end{document}